\documentclass[11pt,a4paper]{article}
\usepackage{jcappub}
\usepackage{bm}
\usepackage{epsfig}
\usepackage{bbold}
\usepackage{graphicx}
\usepackage{amsmath}
\usepackage{amssymb}
\usepackage{amsbsy}
\usepackage{color}
\usepackage{subfigure}
\usepackage{slashed}
\usepackage{afterpage}
\usepackage{psfrag}
\usepackage{hyperref}
\usepackage{dsfont}
\newcommand{\appropto}{\mathrel{\vcenter{
  \offinterlineskip\halign{\hfil$##$\cr
    \propto\cr\noalign{\kern2pt}\sim\cr\noalign{\kern-2pt}}}}}

\renewcommand\vec[1]{{\bf #1}}

\newcommand{\es}[2] {\begin{equation} \label{#1} \begin{split} #2 \end{split} \end{equation}}
\newcommand{\be}{\begin{equation}}
\newcommand{\ee}{\end{equation}}
\newcommand{\bea}{\begin{eqnarray}}
\newcommand{\eea}{\end{eqnarray}}

\newcommand{\exclude}[1]{}

\title{Axion-photon conversion in strongly magnetised plasmas}
\author[a,b]{Alexander J. Millar,}
\author[c]{Sebastian Baum,}
\author[a,b]{Matthew Lawson,}
 \author[a]{M.C. David~Marsh}

\emailAdd{alexander.millar@fysik.su.se}
\emailAdd{sbaum@stanford.edu}
\emailAdd{matthew.lawson@fysik.su.se}
\emailAdd{david.marsh@fysik.su.se}

\affiliation[a]{The Oskar Klein Centre for Cosmoparticle Physics,
Department of Physics,
Stockholm University, AlbaNova, 10691 Stockholm, Sweden}

\affiliation[b]{Nordita, KTH Royal Institute of Technology and
Stockholm
  University, Hannes Alfvéns väg 12, SE-106 91 Stockholm, Sweden}

\affiliation[c]{Stanford Institute for Theoretical Physics, Stanford University, Stanford, CA~94305, USA}

\begin{document}
\subheader{\hfill NORDITA 2021-064}

\abstract{
Axion dark matter can resonantly convert to photons in the magnetosphere of neutron stars, possibly giving rise to radio signals observable on Earth. This method for the indirect detection of axion dark matter has recently received significant attention in the literature. The calculation of the radio signal is complicated by a number of effects; most importantly, the gravitational infall of the axions onto the neutron star accelerates them to semi-relativistic speed, and the neutron star magnetosphere is highly anisotropic. Both of these factors complicate the calculation of the conversion of axions to photons. In this work, we present the first fully three-dimensional calculation of the axion-photon conversion in highly magnetised anisotropic media. Depending on the axion trajectory, this calculation leads to orders-of-magnitude differences in the conversion compared to the simplified one-dimensional calculation used so far in the literature, altering the directionality of the produced photons. Our results will have important implications for the radio signal one would observe in a telescope. 
}

\maketitle

\section{Introduction}

Axions comprise a class of hypothetical particles that were first conceived as a consequence of the `Peccei-Quinn' solution to the strong CP-problem in particle physics~\cite{Peccei:1977hh, Peccei:1977ur,Weinberg:1977ma, Wilczek:1977pj}. Axions are naturally produced in the early universe, and provide an increasingly popular candidate for explaining dark matter~\cite{Preskill:1982cy, Abbott:1982af, Dine:1982ah, Duffy:2009ig, Arias:2012az, Irastorza:2018dyq, Semertzidis:2021rxs, Chadha-Day:2021szb}.\footnote{For the purposes of this work, we indiscriminately refer to both QCD axions and axion-like particles (ALPs) when we write `axion'. For our numerical results, we will assume a relation between the axion mass and the axion-photon coupling as appropriate for a QCD axion, but in all of our equations we keep the axion mass and its couplings as independent parameters allowing to apply our results to both QCD axions and ALPs.} Testing the axion dark matter hypothesis is a central goal of contemporary fundamental physics. 

Any search for axion dark matter relies on the transfer of energy from the astrophysical axions into Standard Model particles. One promising route to achieve this is through the axion-photon coupling, which receives `universal' contributions from axion-pion mixing, thus providing a rather robust and model-independent target. Most experimental and observational ideas to search for axion dark matter are based on the axion-photon coupling.

Astronomical searches for axion dark matter via the axion-photon coupling seek to leverage strong astrophysical magnetic fields to effectively transfer energy from the axion field into photons (or vice versa). Particularly interesting are environments where plasma effects generate an effective mass for the photon that allows the axion and photon dispersion relations to intersect. In such a degenerate situation, dark matter axions can {\it resonantly} convert into photons, enhancing the photon signal and possibly leading to signals detectable with telescopes on Earth. 

Neutron stars are ideal astrophysical environments for converting dark matter axions into photons~\cite{Pshirkov:2007st,Hook:2018iia}: Neutron stars can host extremely strong magnetic fields (up to $\sim 10^{15}$\,G), and are surrounded by a plasma that admits resonant axion-photon conversion near the surface of the star. Due to the plasma densities in the neutron star magnetospheres, resonant conversion is most promising for axions in the neV--meV mass range, including the perhaps best motivated $\mu$eV--meV mass range for QCD axion dark matter. The corresponding photon signal would be a narrow spectral line in the MHz--THz (radio) frequency range, with its frequency set by the axion mass. Radio telescopes combine excellent spatial and frequency resolution with sensitivity to very weak radio signals. For these reasons, axion-photon conversion in neutron star magnetospheres has recently become a very active topic of research, and significant effort has been devoted to characterising the possible signal~\cite{Lai:2006af,Huang:2018lxq,Safdi:2018oeu,Edwards:2019tzf,Battye:2019aco,Leroy:2019ghm,Buckley:2020fmh,Nurmi:2021xds,Witte:2021arp,Battye:2021xvt}, and searching for it in radio data~\cite{Foster:2020pgt,Darling:2020uyo,Darling:2020plz,Battye:2021yue}.

In order to calculate the radio signal from the conversion of dark matter axions into photons in the magnetosphere of neutron stars, one must model a number of effects. First, the phase space of the axion dark matter evolves under the gravitational influence of the neutron star. Second, in the region where the effective photon mass approximately matches the axion mass, one needs to solve the system of equations describing the axion-electrodynamics to compute axion-photon conversion. Finally, one must propagate the electromagnetic modes excited by the axion through the inhomogeneous magnetosphere to the observer far away from the neutron star. Regarding the first effect, previous work has either assumed that axions are on purely radial orbits (in the frame on the neutron star) as they cross the zone of resonant conversion, or that the axion velocity distribution is isotropic at the conversion zone; neither of these assumptions is correct. We discuss how to compute the appropriate axion phase-space density under the influence of the gravitational field of the neutron star for a neutron star with speed relative to the rest frame of the dark matter distribution comparable to the dark matter velocity dispersion, as one expects for a realistic neutron star. The main work in this paper is to present a calculation of the resonant axion-photon conversion in the neutron star magnetosphere (i.e., the second effect above) which fully accounts for the anisotropic plasma density in the magnetosphere. This fully three-dimensional calculation improves over the simplified one-dimensional calculation previously used in the literature, and we will discuss it in more detail in the following paragraph. We do not perform any calculation of the third effect, the propagation of the photons from the conversion zone to the observer at infinity, in this work. This effect can be dealt with by `ray-tracing' calculations, following the electromagnetic modes through the anisotropic plasma of the neutron star magnetosphere, see Refs.~\cite{Leroy:2019ghm,Witte:2021arp,Battye:2021xvt}.

Let us now discuss the core of this work, the calculation of resonant axion-photon conversion in highly magnetised anisotropic plasmas, e.g., in the magnetosphere of a neutron star. A central assumption of essentially all previous studies of this problem is the use of an effectively one-dimensional formalism for calculating the conversion rate of axions into photons, based on the well known calculation for transverse modes~\cite{Raffelt:1987im}.\footnote{An interesting partial exception is Ref.~\cite{Battye:2019aco}, which acknowledged the limitations of the one-dimensional formalism and developed several results for a two-dimensional setting, but left the full three-dimensional problem for further studies. Similarly, in Ref.~\cite{Witte:2021arp} the change of media in other directions was taken as a limiting factor in the conversion probability without solving the 3D equations.} The one-dimensional, linearised system can be expressed as a `Schr{\"o}dinger-like' equation, with time replaced by the spatial coordinate along the direction of propagation. The `conversion probability' of axions into photons calculated from this equation can then be interpreted as the flux transfer: i.e.~the ratio of energy densities of the transverse mode of the electric field and the axion. However, in this paper we show that the intuition built around transverse modes breaks down in highly aniosotropic media, such as the magnetosphere of a neutron star. The anisotropy causes a strong mixing between different photon polarisations, so that propagating states are no longer purely transverse or longitudinal.\footnote{While this has been noted in~\cite{Witte:2021arp}, the implications for how one calculates axion-photon conversion have not yet been explored.} It follows from this  that the naive `conversion probability' no longer describes the physically interesting flux transfer. 

We revisit the classical problem of axion-photon mixing in anisotropic media, including relativistic plasma effects, and derive a new `Schr{\"o}dinger-like' equation governing the evolution of the Langmuir-O (LO) mode, which is the propagating mode in the plasma that is excited by axion-photon conversion. Our calculation has several important implications. First, the anisotropy of the neutron star magnetosphere results in the photon mode evolving along a different direction than that given by its momentum. This invalidates the one-dimensional analysis used in previous work, and can lead to  orders-of-magnitude change in the axion-photon mixing for certain axion trajectories. Second, the `Schrödinger-like equation' now includes damping/growth terms, that can invalidate the  unitary evolution of the system. Third, the energy stored in the  LO modes is {\it not} given by the transverse portion of the field only, and the full Hamiltonian must be used to calculate the flux transfer. Consequently, the flux transfer is no longer identical to the `conversion probability' calculated from the Schrödinger-like equation, so that a naive over-reliance on the quantum mechanical analogue would lead to incorrect conclusions about axion-photon mixing.   

The remainder of this paper is organised as follows. We begin by discussing axion-photon conversion in an anistropic plasma and a magnetic field in section~\ref{sec:conversion}, and derive the new form of the `Schrödinger-like' equation. In section~\ref{probability}, we find the relation between the `axion-photon conversion probability' and the flux transfer. In section~\ref{phasespace} we discuss the calculation of the axion phase space in the neutron star magnetosphere for an neutron star moving with arbitrary speed relative to the rest frame of the dark matter distribution. In section~\ref{sec:numres} we present numerical results for our conversion calculation and compare them to the results one would obtain in the simpler one-dimensional calculation. As we will see, the conversion probability can differ by orders of magnitude between our calculation that fully accounts for the anisotropy of the plasma and the simplified one-dimensional calculation. We conclude in section~\ref{sec:conclusions}. Further technical details of our calculation of axion-photon conversion can be found in the appendices.

\section{Axion-photon conversion in aniostropic plasmas} \label{sec:conversion}
We consider dark matter axions that are gravitationally accelerated towards a neutron star and may convert to photons in its magnetosphere. Due to the large gravitational field of the neutron star, the axions will generally be mildly relativistic, a particularly complicated regime from the point of view of calculating the axion-photon conversion. In such a regime, medium effects cannot be ignored, however the axion also has significant momentum and so the conversion has a strong directional dependence. Our calculation will focus on the regime where the WKB approximation can be applied, i.e. where the axion has momentum larger than the first derivatives of the $E$-fields. As long as this condition holds, our calculation will apply regardless of the relativistic (or not) nature of the axion. Axion-photon mixing is maximised in regions satisfying the resonance condition, which is approximately given when the plasma frequency is equal to the axion mass $\omega_{p} \simeq m_a$. Since this condition is typically fulfilled only in small regions, we solve the mixing problem in the flat-space approximation, neglecting gravity. This approximation is expected to be accurate for most (but not all) possible trajectories of infalling axions.

The classical theory of axions coupled to electromagnetism is governed by the Klein-Gordon equation and Maxwell's equations, modified to include the axion-photon mixing.  Following common practice (see, for example, Refs.~\cite{Sikivie:1983ip,Millar:2017eoc,Hook:2018iia})  we can linearise the equations of motion in a background with the magnetic field ${\bf B}_{\rm NS}$, and restrict (without loss of generality) to electromagnetic modes that oscillate with frequency $\omega$:  
\begin{align}
	\omega^2a+\nabla^2 a&=m_a^2 a-g_{a\gamma} {\bf E}\cdot {\bf B_{\rm NS}} \label{eq:KG1}\\
	-\nabla^2{\bf E}+\nabla(\nabla\cdot {\bf E})&=\omega^2 {\bf D}+\omega^2 g_{a\gamma} a {\bf B_{\rm NS}}\,, \label{eq:wave1}
\end{align}
where we have set the relative permeability to $\mu=1$. The electric field is denoted by ${\bf E}$, and the displacement field by ${\bf D}$. We would now like to solve these equations for dark matter axions traversing a neutron star magnetosphere.

We  must first express the equations \eqref{eq:KG1} and \eqref{eq:wave1} in a convenient coordinate system for any given axion. Two basis vectors can be defined respectively by the direction of motion, and the external magnetic field, which generically are linearly independent. We denote the axion's direction of motion by $\vec{\hat z}$, so that ${\bf k} = k \vec{\hat z}$. This will also be the initial direction of any produced photons, though the changing medium will result in the photon momentum evolving over longer distance scales. We take ${\bf B}_{\rm NS}$ to lie in the $(y, z)$-plane at an angle $\theta(z)$ with the positive $z$-axis. In the following, we assume $\theta\neq \{0,\pi\}$ unless otherwise specified. This defines $\vec{\hat y}(z)$ and consequently also $\vec{\hat x}(z)$ by $\vec{\hat x}(z)= \vec{\hat y}(z) \times \vec{\hat z}$. We will see that the conversion generally occurs over a small region, and so we will neglect gravity and assume that the axion travels along a straight line and $\tfrac{d}{dz} \vec{\hat z} =0$. It then follows that $\tfrac{d}{dz} \vec{\hat x} \propto \vec{\hat y}$ and $\tfrac{d}{dz} \vec{\hat y} \propto \vec{\hat x}$. The expressions for the differential operators $\nabla^2$ and $\nabla(\nabla \cdot {\bf E})$ in the $(\vec{\hat x}, \vec{\hat y}, \vec{\hat z})$ basis are then identical to those in a fixed Cartesian coordinate system. In general, $\theta$ varies due to gravitational bending, the resulting photon refracting in the medium and the fact that the magnetic field itself changes. Again as the region where axion-photon conversion occurs should be small, we take $\theta$ to be almost constant.

The electric displacement field for a highly magnetised plasma is given by the constitutive relation~\cite{gurevich_beskin_istomin_1993} 
\es{}{
\mathbf{D} = \epsilon_{ij}E^j=\begin{pmatrix}
    1      & 0 & 0  \\
    0  & 1-\left\langle\frac{\omega_p^2}{\gamma^3\tilde\omega^2}\right\rangle\sin^2\theta &\left\langle\frac{\omega_p^2}{\gamma^3\tilde\omega^2}\right\rangle\cos\theta\sin\theta  \\
    0 & \left\langle\frac{\omega_p^2}{\gamma^3\tilde\omega^2}\right\rangle\cos\theta\sin\theta & 1-\left\langle\frac{\omega_p^2}{\gamma^3\tilde\omega^2}\right\rangle\cos^2\theta
\end{pmatrix}  \cdot \begin{pmatrix}
    E_x \\
    E_y   \\
    E_z 
\end{pmatrix},
}
 where $\epsilon_{ij}$ is the dielectric tensor with the plasma frequency $\omega_p$. One can see that the anisotropy of the dielectric tensor mixes $E_y$ and $E_z$, so one is not in general left with purely transverse of longitudinal propagating states. Here, we have assumed that the particles forming the plasma are moving along the field lines with speed $v_{||}$ in the neutron star frame, and define
\begin{equation}
	\tilde\omega=\omega-k_{||}v_{||}\,,\quad \gamma=\frac{1}{\sqrt{1-v_{||}^2}}\,.
\end{equation}
We have also defined 
\begin{equation}
    \langle... \rangle\equiv\sum_{i}\int_{-\infty}^{\infty}dp_{||}...F^i_{||}(p_{||})\,
\end{equation}
where $F^i_{||}$ is the distribution function of the species $i$ in the plasma. Note that for an ultrarelativistic plasma, $\langle\gamma^{-3}\rangle\simeq \langle\gamma\rangle^{-1}$~\cite{PhysRevE.57.3399}.
In Refs.~\cite{Hook:2018iia,Battye:2019aco,Leroy:2019ghm}, the electrons in the plasma were treated as non-relativistic, i.e. $\gamma=1$. While we allow for relativistic electrons, we note that infalling axions will be at most mildly relativistic. For mildly relativistic axions, one would expect $\tilde\omega\simeq\omega$, however, in principle the distinction between $\tilde\omega$ and $\omega$ can lead to differences for high momenta axions traveling through relativistic plasmas. For dark matter falling into a neutron star, this will simply change the effective $\omega_p$ by a factor of $\tilde\omega/\omega$ which is trivial compared to the significant modelling uncertainties of neutron stars (discussed in section \ref{NSmodel}). 
	
When the axions are decoupled from the system, one can solve for the propagating electromagnetic states of the medium. Note that such states only apply locally: as one moves through the medium, the eigenstates deform adiabatically. As noted in~\cite{Witte:2021arp}, the relevant electromagnetic mode (i.e. the mode excited by axion-photon conversions) is the so-called Langmuir-O (LO) mode, whose polarization has both longitudinal and transverse components. As we discuss in Appendix~\ref{longitudinal}, in principle, the axion can also excite purely longitudinal modes. However, such purely longitudinal modes are non-propagating and are hence irrelevant for the electromagnetic signal observed far from the neutron star. The LO mode, on the other hand, evolves adiabatically into a purely transverse vacuum mode as it exits the neutron star atmosphere~\cite{PhysRevE.57.3399}. The specific form of the LO mode is found in Appendix~\ref{longitudinal} to be
\begin{equation}
    {\bf\hat E}_{\rm LO}=\frac{-1}{\sqrt{1+\frac{\bar\omega_p^4\cos^2\theta\sin^2\theta}{(\omega^2-\bar\omega_p^2\cos^2\theta)^2}}}\left({\bf \hat y}-\frac{\bar\omega_p^2\cos\theta\sin\theta}{\omega^2-\bar\omega_p^2\cos^2\theta}{\bf \hat z}\right )\,,\label{eq:LO}
\end{equation}
where we have defined an effective relativistic plasma frequency $\bar\omega_p^2=\omega^2\langle\omega_p^2\gamma^{-3}/\tilde\omega^2\rangle$. In the non-relativistic limit, as has been mostly considered in the literature, there is no distinction between $\bar\omega_p$ and $\omega_p$. Our task is to calculate the conversion of axions into the LO mode. While the modification to the dispersion relation was noted in Ref.~\cite{Witte:2021arp}, the effects of the mixing of transverse and longitudinal modes on the conversion calculation and the implications for how one defines a photon were not explored. As we will see, such effects can change the conversion probability for a given axion trajectory by orders of magnitude.
	
We can now write out the $x,y,z$ components of equation~\eqref{eq:wave1}. We will be considering the case of axion photon mixing in a slowly varying plasma, so we can neglect second order derivatives that do not involve the $z$-direction. In addition, $E_x$  does not directly couple to the axion as, by construction $\hat{x} \bot \vec{B}_{\rm NS}$. While $E_x$ can be generated by derivative terms, it is suppressed when compared to $E_{y,z}$ so we can also neglect it. At lowest order this leaves the following coupled equations:
\begin{align}
	\frac{\partial^2 E_z}{\partial z\partial y}&\simeq(\omega^2-\bar\omega_p^2\sin^2\theta)E_y+\bar\omega_p^2\cos\theta\sin\theta E_z+\frac{\partial^2 E_y}{\partial^2 z}-\omega^2 g_{a\gamma} a B_{\rm NS}\sin\theta\,,
	\label{eq:mixed11}\\
	\frac{\partial^2 E_y}{\partial z\partial y}&\simeq(\omega^2-\bar\omega_p^2\cos^2\theta)E_z+\bar\omega_p^2\cos\theta\sin\theta E_y+\omega^2 g_{a\gamma} a B_{\rm NS}\cos\theta\,.  \label{eq:mixed21}
\end{align}    
Previous studies of axion-photon conversion near neutron stars have all neglected 
the left-hand side of these equations, featuring mixed $y$- and $z$-derivatives.\footnote{While it is noted that the full gradient needs to be dealt with in Ref.~\cite{Battye:2019aco,Witte:2021arp}, the conversion was not calculated including mixing terms. All previous calculations including Ref.~\cite{Hook:2018iia, Leroy:2019ghm} similarly neglect the mixed derivatives.} Neglecting mixed derivatives ignores the mixing between $E_y$ and $E_z$ and is equivalent to assuming that $E_y$ only evolves along $z$. This does not hold for general trajectories, and is only a good approximation if $\theta\simeq \pi/2$, i.e., when the axion happens to be traveling perpendicularly to $\vec{B}_{\rm NS}$. To simplify matters, we can eliminate $E_z$ from the equations. This will allow us to solve directly for the propagating mode, and neglect non-propagating contributions to $E_z$, which are discussed in Appendix~\ref{longitudinal}. By substituting  equation \eqref{eq:mixed21} into \eqref{eq:mixed11} and neglecting any higher order derivatives that follow, we end up with a single partial differential equation for $E_y$ and $a$:
\begin{align}
    &	-\frac{\partial^2 E_y}{\partial^2 z}-\frac{2\bar\omega_p^2\cos\theta\sin\theta}{\omega^2-\bar\omega_p^2\cos^2\theta}\frac{\partial^2 E_y}{\partial z\partial y}-\frac{2\omega^2\bar\omega_p\sin\theta\cos\theta}{(\omega^2-\bar\omega_p^2\cos^2\theta)^2}\frac{\partial \bar\omega_p}{\partial y} 
    \frac{\partial E_y}{\partial z}-\frac{2\omega^2\bar\omega_p\sin\theta\cos\theta}{(\omega^2-\bar\omega_p^2\cos^2\theta)^2}\frac{\partial \bar\omega_p}{\partial z} 
    \frac{\partial E_y}{\partial y}\nonumber\\
    \simeq ~& \frac{\omega^2-\bar\omega_p^2}{1-\frac{\bar\omega_p^2}{\omega^2}\cos^2\theta}E_y  -\frac{\omega^2\sin\theta}{1-\frac{\bar\omega_p^2}{\omega^2}\cos^2\theta} g_{a\gamma} a B_{\rm NS}\,. \label{eq:geney}
\end{align}
We neglect also the derivatives of the axion driving term, as they will only have a small impact on the evolution of $E_y$. 

To attempt an analytic solution, we assume that the change in $\bar\omega_p$, and so $k$, is small and thus the direction of propagation stays the same.  Of course, this is not true in general: photon trajectories will bend in a changing medium. Assuming small changes in $\bar\omega_p$, $k$ necessarily restricts us to solving over some finite region where the photon path is approximately constant. The region over which the majority of the generation of an axion-induced $E$-field occurs is referred to as the conversion zone. While one can estimate the radius of curvature for the changing direction of the photon, such as was done in Ref.~\cite{Witte:2021arp}, our focus in this work is not on the propagation of the converted photons. As most of the generation of the propagating LO mode via the axion should be localised to a smaller region, separating propagation and conversion allows for an analytic approximation of the fields generated by the axion, which could then be propagated numerically via ray tracing code such as in Refs.~\cite{Witte:2021arp,Battye:2021xvt}.

As long as we are treating the medium as slowly varying, we can write the fields as an envelope term, denoted by a tilde, multiplying a plane wave,
\begin{equation}
	E_y\equiv\tilde E_y(y,z)e^{i(\omega t-k z)}\,,\quad\quad\quad a\equiv \tilde a (z)e^{i(\omega t-k z)}\,. 
\end{equation}  
To move further, we will take the WKB approximation and assume that $k\tilde E_y(y,z)\gg \partial \tilde E_y(y,z)/\partial y,\partial \tilde E_y(y,z)/\partial z\gg \partial^2 \tilde E_y(y,z)/\partial z^2$. As we will see, the smaller the derivatives the longer the length scales over which axions can resonantly convert, leading to higher conversion probabilities. Thus a breakdown of the WKB approximation, which occurs when the first derivatives of the $\bar\omega_p$ vanish, corresponds to a region of enhanced axion-photon conversion. The WKB approximation allows us to then simplify Eq.~\eqref{eq:geney} to 
\begin{equation}
	2ik\frac{\partial \tilde E_y}{\partial z}+2ik \frac{\bar\omega_p^2 \xi}{\omega^2\tan\theta} \frac{ \partial\tilde E_y}{\partial y}\simeq(m_a^2-\xi \bar\omega_p^2-ik{\cal D})\tilde E_y  -\frac{\omega^2 \xi}{\sin\theta} g_{a\gamma} \tilde a B_{\rm NS}\,, \label{eq:linearised}
\end{equation}
where we have used the on-shell relation for the axion, $m_a^2=\omega^2-k^2$, and defined
\begin{equation}
	\xi\equiv\frac{\sin^2\theta}{1-\frac{\bar\omega_p^2}{\omega^2}\cos^2\theta}\,.
\end{equation} 
The damping/growth term ${\cal D}$ appears as an artifact of assuming that the photon does not experience any curvature, i.e., that it is well described as only having momentum in the $z$-direction, and is given by 
\begin{equation}
    {\cal D}\equiv \frac{2\omega_p\xi^2}{\omega^2\tan\theta\sin^2\theta}\frac{\partial{\bar \omega_p}}{\partial y}\,.
\end{equation}
The back-reaction on to the axion field should be small, so we can treat $\tilde  a$ as constant. 

\begin{figure*}
    \centering
    \includegraphics[trim = 0mm 0mm 0mm 0mm, clip, width=.4\textwidth]{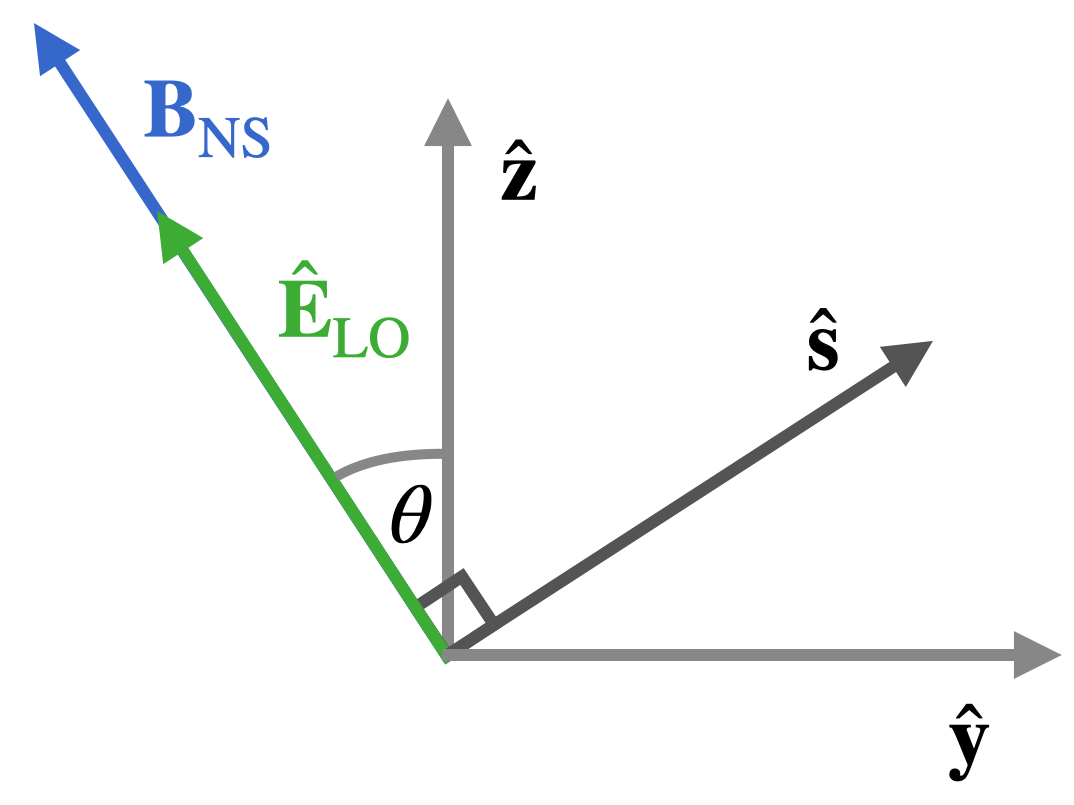} 
    \caption{We show the various axes associated with axion-photon conversion in an anisotropic medium. The axion propagates in the $\bf \hat z$ direction, which is at an angle $\theta$ from the magnetic field ${\bf \hat B}_{\rm NS}$. The magnetic field is defined to lie in the ${\bf \hat z},{\bf \hat y}$ plane. In the non-relativistic limit, the $E$-field associated with the propagating LO mode, ${\bf \hat E}_{\rm LO}$, is given by ${\bf \hat B}_{\rm NS}$, i.e., ${\bf \hat E}_{\rm LO}$ is parallel to the $B$-field. The magnitude of ${\bf \hat E}_{\rm LO}$ evolves along the perpendicular direction, denoted by $\bf \hat s$. }
 	\label{fig:axes}
\end{figure*}

To progress, we can write a new differential operator
\begin{equation}
    \frac{\partial}{\partial s}=\frac{\partial}{\partial z}+\frac{\bar\omega_p^2 \xi}{\omega^2\tan\theta}\frac{\partial}{\partial y}\,.
\end{equation}
This operator allows us to finally write the axion-photon mixing equations as a Schr\"odinger-like equation
\begin{align}
   i \frac{\partial \tilde E_y}{\partial s}=\frac{1}{2k}\left (m_a^2-\xi\bar\omega_p^2-ik{\cal D}\right )\tilde E_y -\frac{1}{2k}\frac{\omega^2 \xi}{\sin\theta} g_{a\gamma} \tilde a B_{\rm NS}\,, \label{eq:schro}
\end{align}
As $\partial/\partial s$ is a function of $\bar\omega_p$ and $\theta$, and in general $\bar\omega_p(s)$, any integration over $s$ will in principle be curvilinear. However, if we are only concerned with small regions, we can assume that $\bar\omega_p$ is only slowly varying and so $s$ is constant with respect to $y, z$. Assuming a constant $\bar\omega_p$ we can write
\begin{equation}
    {\bf \hat   s}\simeq\frac{1}{\sqrt{1+\frac{\bar\omega_p^4\cos^2\theta\sin^2\theta}{(\omega^2-\bar\omega_p^2\cos^2\theta)^2}}}\left({\frac{\bar\omega_p^2\cos\theta\sin\theta}{\omega^2-\bar\omega_p^2\cos^2\theta}{\bf \hat y}+\bf \hat z}\right)\,.
\end{equation}
Comparing with Eq.~\eqref{eq:LO} we see that ${\bf \hat s}$ is actually perpendicular to ${\bf \hat E}_{\rm LO}$. In other words, the propagating mode $E_{\rm LO}$ is transverse with respect to the direction it evolves in, rather than the direction of propagation. While the mode is not transverse, it evolves in a similar manner to a transverse wave. Note that $\partial /\partial s$ is not defined with unit norm for simplicity, however that means one has to be slightly careful when talking about length scales associated with $s$. For non-relativistic axions ($\bar \omega_p=m_a$), we can write down simple expressions
\begin{subequations}
\begin{align}
    {\bf \hat B}_{\rm NS} &= -\sin\theta\, {\bf \hat y}+\cos\theta\,{\bf \hat z}\,,\\
    {\bf \hat E}_{\rm LO} &\simeq -\sin\theta\, {\bf \hat y}+\cos\theta\,{\bf \hat z}\,,\\
    {\bf \hat s} &\simeq \cos\theta\, {\bf \hat y}+\sin\theta\,{\bf \hat z}   \,.
\end{align}
\end{subequations}
In this limit we can simply show the relationship between $B_{\rm NS}$, $z$, $E_{\rm LO}$ and $s$, depicted in Fig.~\ref{fig:axes}. We can see in this limit that ${\vec B}_{\rm NS}||{\vec E}_{\rm LO}$: the LO mode couples perfectly to the axion via ${\vec E} \cdot {\vec B}$. For ultrarelativsitic axions with $\omega\gg\bar\omega_p$, we see instead $\vec{ \hat E}_{\rm}\simeq -\vec{ \hat y}$, $\vec{\hat s}\simeq\vec{\hat z}$ and $\xi\simeq\sin^2\theta$ giving back the usual mixing of an axion with a transverse photon~\cite{Raffelt:1987im}.

Ignoring the overall phase, as shown in Appendix~\ref{stationaryphase}, Eq.~\eqref{eq:schro} can be solved in the stationary phase approximation to give
\begin{equation}
    \tilde E_y[s_0+L/2]=\omega^2 \sqrt{\frac{\pi}{2k\left |\bar\omega_p\bar\omega_p'+\frac{\omega^2-\bar\omega_p^2}{\omega^2\tan\theta}\bar\omega_p^2\theta'\right |}}e^{\int_0^{s_0}ds'{\cal D}/2}g_{a\gamma} \tilde a B_{\rm NS}\,,\label{eq:stationary}
\end{equation}
where $\bar\omega_p'=\partial \bar\omega_p/\partial s$, we have allowed for a small change in $\theta$ given by $\theta'=\partial \theta/\partial s$ and $s_0$ is given by solving for the resonance condition
\begin{equation}
	\bar\omega_p(s)^2=\frac{m_a^2\omega^2}{m_a^2\cos^2\theta+\omega^2\sin^2\theta}\,.\label{eq:resonancecond}
\end{equation}
One can see how the LO mode extrapolates between longitudinal and transverse behaviour as one varies $\theta$. If $\theta=\pi/2$, the dispersion relation is that of a regular transverse mode, with a resonance given by $\bar\omega_p=m_a$. If instead $\theta=0$, then the resonance condition is that for an axion converting to a longitudinal plasmon with negligible spatial dispersion, $\bar\omega_p=\omega$, as for example studied in Ref.~\cite{Caputo:2020quz}. In the non-relativistic limit, $\omega \to m_a$, and the resonance condition simply becomes $\bar\omega_p = m_a$ for any $\theta$. The conversion length can be read off from the exponent in the stationary phase approximation (similar to Ref.~\cite{Battye:2019aco}), giving
\begin{equation}
    L=\frac{\sin \theta}{\xi}\sqrt{\frac{\pi k}{\left |\bar\omega_p\bar\omega_p'+\frac{\omega^2-\bar\omega_p^2}{\omega^2\tan\theta}\bar\omega_p^2\theta'\right |}}\simeq\sin\theta\sqrt{\frac{\pi k}{\bar\omega_p|\bar\omega_p'|}}\,,
\end{equation}
where the latter statement holds in the non-relativistic limit.
To evaluate the damping term in Eq.~\eqref{eq:stationary}, we note that by assumption, the equation is only valid in a small region around the conversion zone where higher order derivatives and the bending of photons can be neglected. Assuming that $\partial \bar\omega_p/\partial y={\cal O}(\partial\bar \omega_p/\partial s)$, one can see that over the conversion zone ($s_0\pm L/2$)
\begin{equation}
    \int_{s_0-L/2}^{s_0+L/2}ds'{\cal D}={\cal O}\left(\frac{k}{m_a^2L}\right)\ll 1\,,
\end{equation}
thus, for simplicity, we will neglect this damping in the probability of conversion.
We can also include the longitudinal component of the LO mode, giving
\begin{align}
\tilde E_y[s_0+L/2]&\simeq	g_{a\gamma}\tilde a B_{\rm NS}\omega^2\sqrt{\frac{\pi  }{2 k \left |\bar\omega_p\bar\omega_p'+\frac{\omega^2-\bar\omega_p^2}{\omega^2\tan\theta}\bar\omega_p^2\theta'\right |}} \,,\label{eq:partial1}\\ 
\tilde E_z[s_0+L/2]&\simeq	g_{a\gamma}\tilde a B_{\rm NS}\omega^2\frac{\bar\omega_p^2\cos\theta\sin\theta}{\omega^2-\bar\omega_p^2\cos^2\theta}\sqrt{\frac{\pi   }{2 k \left |\bar\omega_p\bar\omega_p'+\frac{\omega^2-\bar\omega_p^2}{\omega^2\tan\theta}\bar\omega_p^2\theta'\right |}}\,,\label{eq:efull}	
\end{align}
Actually, as discussed in Appendix~\ref{longitudinal}, there are two contributions to $E_z$. As we are concerned with propagating modes that may exit the neutron star, here we only consider the longitudinal component of the LO mode. There is also a term that directly comes from the axion-$E_z$ mixing, however, it only persists in the region where the axion mixing is strong, and follows the phase of the axion, rather than the LO mode. Thus, after a length $L$, this axion-mixing term will dephase with the LO mode and not contribute to the $E$-fields exiting the neutron star. 
If we then take $\bar\omega_p=\omega=m_a$, we get the simple expressions
\begin{align}
\tilde E_y[s_0+L/2]&\simeq	g_{a\gamma}\tilde a B_{\rm NS} \sqrt{\frac{\pi m_a^3}{2k|\bar\omega_p'|}}\,,\\
\tilde E_z[s_0+L/2]&\simeq	g_{a\gamma}\tilde a B_{\rm NS} \frac{m_a^2\cos\theta \sin \theta}{k^2+m_a^2\sin^2\theta}\sqrt{\frac{\pi m_a^3}{2k|\bar\omega_p'|}}\,.\label{eq:esimple}
\end{align}
 We kept the factor of $k^2$ in the denominator of $E_z$ to avoid an unphysical divergence at $\theta\to \{0,\pi\}$. Note that the $E_z$ component can actually dominate the $E$-field:
 \begin{equation}
     \frac{E_z}{E_y}=\frac{m_a^2\cos\theta \sin \theta}{k^2+m_a^2\sin^2\theta}\,,
 \end{equation}
 which has a maximum at $|E_z/E_y|\simeq m_a/2k$ for $k\ll m_a$. Even for large $k$, $E_y$ and $E_z$ are generically of similar magnitude.

\section{Probability of conversion} \label{probability}
When expressing the classical problem of axion-photon mixing as a Schrödinger-like equation (with time replaced by the spatial coordinate $z$ or $s$), it is natural to make use of the language and techniques originally developed within quantum mechanics to simplify calculations. The most natural object to compute from the Schrödinger equation is the `conversion probability' ($|A_y/a|^2$), i.e.~the norm square of the transition amplitude of going from a pure axion state ($a=1$) into a pure photon state ($A_y$). However, it is important to remember that this is not actually a quantum mechanical calculation, for which one would have to use the correct Schrödinger equation (with time derivatives and $\hbar \neq 0$), and carefully defined wavefunctions. In the vaccuum case, the `conversion probability' calculated from the classical Schrödinger-like equation is of direct physical interest as it gives the ratio of the energy densities stored in axion and electromagnetic fields~\cite{Raffelt:1987im}; this definition of conversion probability has been used in Refs.~\cite{Hook:2018iia,Battye:2019aco,Leroy:2019ghm,Witte:2021arp} in the context of axion-photon conversion near neutron stars. However, as we discuss in this section, when taking into account the 3D complexity of the neutron star magnetosphere and the correct definition of propagating modes in the plasma, the naive `conversion probability' is no longer the physically relevant object to compute. In essence, such a definition does not include the fact that the Hamiltonian is modified by the presence of matter, so would miscount the number of photons generated if interpreted as an actual conversion probability. 

The key point of this section is that the (time-averaged) energy density stored in the propagating LO mode is no longer simply $1/4(1+\bar\omega_p^2/\omega^2)|E_y|^2$, as is the case for an isotropic non-relativistic plasma. To calculate the energy stored in the LO mode, we must consider that the medium has both temporal dispersion, and is anisotropic, giving~\cite{Bers1983SpacetimeEO,Bers2}
\begin{equation}
	U=\frac{1}{4}\frac{\partial}{\partial \omega}(\omega\epsilon^{ij})E_i^* E_j+\frac{1}{4}H_i^*H^j\,.
\end{equation}
While the $H$-field terms are suppressed at order $k^2/\omega^2$ relative to the $E$-field terms, near the neutron star surface, $k$ can be relatively large compared to $\omega$, meaning that significant energy can be stored in the $H$-fields. For a propagating state given by $E_{\rm LO}$, we find
\begin{equation}
	U=|E_y|^2 \frac{\left(2 \omega^4-\omega^2
   \bar\omega_p^2+\left(\bar\omega_p^4-3 \omega^2
   \bar\omega_p^2\right) \cos (2
   \theta)+\bar\omega_p^4\right)}{8
   \left(\omega^2-\bar\omega_p^2 \cos
   ^2\theta\right)^2}+|E_y|^2 \frac{k^2}{4\omega^2}\,.\label{eq:energy1}
\end{equation}
Equation \eqref{eq:energy1} reduces to $1/4(1+\bar\omega_p^2/\omega^2)|E_y|^2$ in the non-relativistic limit and when $\theta=\pi/2$, and reduces further  to $|E_y|^2/2$ for $\omega=\bar\omega_p$, and it is only in this case that the naive conversion probability ($|A_y/a|^2$) gives the ratio of energy densities stored in the electromagnetic and axion fields. Notably, the energy stored in the propagating mode is enhanced for $B$-fields almost aligned with the direction of propagation. In the absence of medium losses, the energy stored in the photon field should be conserved: when propagated through a slowly changing medium the propagation state will adiabatically deform, but maintain stored energy. Thus, taking only the transverse component of the stored energy into account, as done in previous studies, neglects a significant part of the stored energy, or in other words, miscounts the number of photons that are converted. 

For a simple expression, we can finally write the ratio of the energy densities of the electromagnetic propagating mode and the axion field in the non-relativistic limit ($m_a\gg k$) as
\begin{align}
	R&=\frac{2U}{\omega^2|\tilde a|^2}=\left |\frac{A_y^2}{a^2}\right |\left[ \frac{\left(2 \omega^4-\omega^2
   \bar\omega_p^2+\left(\bar\omega_p^4-3 \omega^2
   \bar\omega_p^2\right) \cos (2
   \theta)+\bar\omega_p^4\right)}{4
   \left(\omega^2-\bar\omega_p^2 \cos
   ^2\theta\right)^2}+\frac{k^2}{2\omega^2}\right ]\nonumber \\
	&\simeq \frac{g_{a\gamma}^2B_{\rm NS}^2}{2k|\bar\omega_p'|}\frac{\pi m_a^5}{(k^2+m_a^2\sin^2\theta)^2}\sin^2\theta\,. \label{eq:prob1}
\end{align}
As the system is considered here to be time independent, a single axion should convert to a single photon with energy $\omega$ and so the ratio of energy densities should correspond to the ratio between the axion and photon fluxes. Note that this \emph{flux transfer} is distinct from the `conversion probability' ($|A_y/a|^2$) calculated from the Schrödinger-like Eq.~\eqref{eq:schro}, essentially reweighting it via the Hamilitonian. However, if one did a full quantum mechanical calculation, the expectation value of the conversion probability of axions to LO photons should agree with this flux transfer~\cite{Raffelt:1991ck}.

Unlike the result of Ref.~\cite{Hook:2018iia}, no propagating photon is generated from a non-relativistic axion in a longitudinal $B$-field (i.e., for $\vec{B}_{\rm NS} \| \hat{z}$). However, we should stress that $\theta=\{0,\pi\}$ is a special condition and the order of limits matters. For purely longitudinal conversion one should perform a dedicated calculation (i.e., solve Eq. \eqref{eq:mixed21} directly or perform an analysis in the vein of Ref.~\cite{Caputo:2020quz}). We should note that Eq.~\eqref{eq:prob1} breaks down under more extreme conditions. If $|\bar\omega_p'|=0$, in other words, if $\bar\omega_p$ is constant along $s$, the conversion length becomes infinite. To regulate such divergences, one must cut-off $L$ when it is competitive compared to other limiting scales, such as the radius of curvature of the photon trajectory or that over which second derivatives are relevant~\cite{Witte:2021arp}, cf.~Appendix~\ref{sec:statphase}. 
 
Our ratio can be contrasted with the `conversion probability' calculated when the non-trivial Hamiltonian and the mixing of transverse and longitudinal modes are neglected,
\begin{equation}
	P^{\rm 1D}_{a\to\gamma}=\frac{\pi\bar\omega_p}{2k|\partial\bar\omega_p/\partial z|}g_{a\gamma}^2B_{\rm NS}^2\sin^2\theta\,.\label{eq:prob1d}
\end{equation} 
If we take $\theta=\pi/2$ so that $\hat{\vec{s}}=\hat {\vec{z}}$, reducing back to a purely one dimensional problem, then we see that both expressions agree, as one would expect
 \begin{equation}
	R\to P^{\rm 1D}_{a\to\gamma}\left(\theta=\frac{\pi}{2}\right) =\frac{\pi\bar\omega_p}{2k|\bar\omega_p'|}g_{a\gamma}^2B_{\rm NS}^2\,.
\end{equation} 
We have now derived in Eq.~\eqref{eq:prob1} an analytic expression for axion-photon conversion in highly aniosotropic systems. In cases where the axion is traveling at a non-trivial angle to the external $B$-field there can be significant alterations to the flux transfer. To see what impact the 3D calculation has in a neutron star, we must consider a specific neutron star model.

\section{Phase space distribution}
\label{phasespace}
In this section, we discuss the phase space distribution of the axions at the conversion surface. An excellent discussion of the change of the phase space distribution of collisionless particles under the influence of the gravitational field of an astrophysical object such as a neutron star can be found in Ref.~\cite{Alenazi:2006wu}. We will assume that far away from the neutron star, the velocity distribution of the axions in the galactic rest frame is given by a Maxwell-Boltzmann distribution with velocity dispersion $\sigma_v$ truncated at the Galactic escape velocity $v_{\rm esc}$ as in the Standard Halo Model~\cite{Drukier:1986tm,Lewin:1995rx,Freese:2012xd},
\begin{equation}
    \tilde{f}_{\vec{v}}^\infty(\vec{v}) = \frac{1}{N_{\rm esc}} \left( \frac{1}{2\pi \sigma_v^2} \right)^\frac{3}{2} e^{-\frac{v^2}{2\sigma_v^2}} H\left( v_{\rm esc} - v \right) \;,
\end{equation}
where $H(x)$ is the Heaviside step function, and the normalization factor is given by
\begin{equation}
    N_{\rm esc} = {\rm Err}\left( \frac{v_{\rm esc}}{\sqrt{2} \sigma_v} \right) - \sqrt{\frac{2}{\pi}} \frac{v_{\rm esc}}{\sigma_v} e^{-\frac{v_{\rm esc}^2}{2 \sigma_v^2}} \;.
\end{equation}
We can shift this distribution into the frame of the neutron star, 
\begin{equation} \label{eq:f3vNS}
    f_{\vec{v}}^\infty(\vec{v}) = \tilde{f}_{\vec{v}}^\infty(\vec{v} - \vec{v}_{\rm NS}) \;,
\end{equation}
where $\vec{v}_{\rm NS}$ is the velocity of the neutron star relative to the galactic rest frame. The six-dimensional phase space density far away from the neutron star is then
\begin{equation} \label{eq:f6NS}
    f_6^\infty(\vec{v}) = n_a^\infty f_{\vec{v}}^\infty(\vec{v}) = \frac{\rho_a^\infty}{m_a} f_{\vec{v}}^\infty(\vec{v}) \;,
\end{equation}
where $n_a^\infty$ ($\rho_a^\infty$) are the axion number (mass) density far away from the neutron star. 

Liouville's theorem states that the phase space density is conserved along trajectories. Hence, at some point $\vec{r}_{\rm NS}$ in spherical coordinates centered on the neutron star, the phase space density is given by
\begin{equation} \label{eq:f6NSLie}
    f_6(\vec{r}_{\rm NS}, \vec{v}) = f_6^\infty(\vec{v}_\infty) \;,
\end{equation}
where $\vec{v}_\infty = \vec{v}_\infty(\vec{r}_{\rm NS}, \vec{v})$ is the velocity at infinity for an orbit with velocity $\vec{v}$ at $\vec{r}_{\rm NS}$. To distinguish from our axion-centric frame as used in the previous sections, we use NS as a subscript; in other words, the position vector is given by ${\bf \hat r}_{\rm NS}=(r_{\rm NS},\theta_{\rm NS},\phi_{\rm NS})$ in spherical coordinates. Let us stress here that Liouville's theorem states only that the phase space {\it density} is conserved along trajectories, it does not make any immediate statements about the relation of the functional form of the phase space density at different $\vec{r}$. In order to obtain that functional form, one must compute $\vec{v}_\infty(\vec{r}_{\rm NS}, \vec{v})$. For a Newtonian potential, energy, angular momentum, and the Laplace-Runge-Lenz vector of a particle moving in the potential are conserved, such that~\cite{Alenazi:2006wu}
\begin{equation} \label{eq:vinf}
    \vec{v}_\infty(\vec{r}_{\rm NS}, \vec{v}) = \frac{ v_\infty^2 \vec{v} + v_\infty \left( G M_{\rm NS}/r_{\rm NS} \right) \hat{\vec{r}}_{\rm NS} - v_\infty \vec{v} \left( \vec{v} \cdot \vec{r}_{\rm NS} \right) }{ v_\infty^2 + \left( G M_{\rm NS} / r_{\rm NS} \right) - v_\infty \left( \vec{v} \cdot \vec{r}_{\rm NS} \right) } \;,
\end{equation}
with $v_\infty = \sqrt{v^2 - 2 G M_{\rm NS} / r_{\rm NS}}$, the gravitational constant $G$, the mass of the neutron star $M_{\rm NS}$, and $\hat{\vec{r}}_{\rm NS} = \vec{r}_{\rm NS}/r_{\rm NS}$.

It is interesting to note that due to the symmetry of the system, if the velocity and mass distribution of the axions is homogeneous and isotropic (in the frame of the neutron star) far away from the neutron star, i.e. $\vec{v}_{\rm NS} = 0$ in Eq.~\eqref{eq:f3vNS}, the velocity distribution will be isotropic at any $\vec{r}_{\rm NS}$. In this case, one can immediately compute~\cite{Leroy:2019ghm}
\begin{equation} \label{eq:Amsterdamf6}
    f_6(\vec{r}_{\rm NS}, \vec{v}) = \frac{\rho_a^\infty}{m_a} \frac{1}{N_{\rm esc}} \left( \frac{1}{2\pi \sigma_v^2} \right)^2 e^{\frac{G M_{\rm NS}}{r_{\rm NS} \sigma_v^2}} e^{-\frac{v^2}{2\sigma_v^2}} H\left( \sqrt{v^2 - 2 G M_{\rm NS} / r_{\rm NS}} - v_{\rm esc} \right) \;.
\end{equation}

With Eqs.~\eqref{eq:f6NS}--\eqref{eq:vinf}, the dark matter number density at $\vec{r}_{\rm NS}$ can be obtained by integrating over all kinematically allowed velocities, $v \gtrsim \sqrt{2 G M_{\rm NS}/r_{\rm NS}}$
\begin{equation}
    n_a(\vec{r}_{\rm NS}) = \frac{\rho_a(\vec{r}_{\rm NS})}{m_a} = \int_{v > \sqrt{2 G M_{\rm NS}/r_{\rm NS}}} d^3v \, f_6(\vec{r}_{\rm NS}, {\vec v}) \;.
\end{equation}
For an homogeneous and isotropic (in the frame of the neutron star) axion phase space distribution far away from the neutron star and $v_{\rm esc} \gg \sigma_v$, one can find an approximate closed solution for $n_a(\vec{r}_{\rm NS})$, see Ref.~\cite{Leroy:2019ghm}. We stress that their solution [and Eq.~\eqref{eq:Amsterdamf6}] is only applicable if, far away from the neutron star, both the axion density and velocity distributions are both homogeneous and isotropic (in the frame of the neutron star) and if $v_{\rm esc} \gg \sigma_v$. A typical neutron star moves with speed $v_{\rm NS}$ of order $\sigma_v$ relative to the Galactic rest frame, breaking the assumption of an isotropic velocity distribution in the frame of the neutron star.

While the full six dimensional phase space contains all relevant information, for any realistic conversion location, $\sqrt{2G M_{\rm NS}/r_{\rm NS}} \gg \{\sigma_v,v_{\rm esc} \}$. Hence, almost the entirety of the axion's speed comes from the infall. Thus, the distribution over $v$ can approximately be treated as a delta-function $\delta(v-\sqrt{2G M_{\rm NS}/r_{\rm NS}})$. In other words, we care about the distribution of incoming angles, rather than the distribution over incoming speeds. This allows us to define
\begin{equation}
    f_5({\bf r}_{\rm NS},\theta_v,\phi_v)=\frac{f_6(\vec{r}_{\rm NS},\sqrt{2G M_{\rm NS}/r_{\rm NS}},\theta_v,\phi_v)}{\int d\Omega_v'f_6(\vec{r}_{\rm NS}, \sqrt{2G M_{\rm NS}/r_{\rm NS}},\theta_v',\phi_v')}\,.\label{eq:fivespace}
\end{equation}
Thus armed, we can calculate the flux transfer for general axion distributions, including non-vanishing speeds of the neutron star relative to the galaxy.  

\section{Comparison with 1D calculations in the Goldreich-Julian Model} \label{sec:numres}
It is clear that for specific values $\theta$ there can be significant variation between the 3D and 1D versions of the axion-photon conversion. As the plasma is generated and shaped by the neutron star magnetic field, the changes in $\omega_p$ are driven by the changing $B$-field, and so may also be correlated with $\theta$. Here we explore what influence a more complete axion-photon conversion calculation has under realistic conditions. After a brief discussion of the Goldreich-Julian neutron star model in section~\ref{NSmodel}, we will compare results for the axion-photon conversion in our full 3D calculation with results in the conventional 1D calculation in sections~\ref{sec:results_direction}--\ref{sec:results_power}.

\subsection{The Goldreich-Julian Model}
\label{NSmodel}
To consider a realistic scenario, we turn to an analytic description of a neutron star. One commonly used model is the Goldreich-Julian (GJ) model~\cite{Goldreich:1969sb}. The GJ model gives a simple analytic solution by requiring a charge density to satisfy the condition ${\bf E}\cdot {\bf B}=0$ at the surface of the neutron star in the presence of a strong, rotating dipole magnetic field given by 
\begin{align}
    B_{r_{\rm NS}} &= B_0 \left(\frac{r_0}{r_{\rm NS}}\right)^3 \left[\cos\theta_m\cos\theta_{\rm NS} + \sin\theta_m\sin\theta_{\rm NS}\cos(\Omega t)\right]\,, \nonumber  \\
    B_{\theta_{\rm NS}} &= \frac{B_0}{2} \left(\frac{r_0}{r_{\rm NS}}\right)^3 \left[\cos\theta_m\sin\theta_{\rm NS} - \sin\theta_m\cos\theta_{\rm NS}\cos(\Omega t)\right]\,, \nonumber \\
    B_{\phi_{\rm NS}} &= \frac{B_0}{2} \left(\frac{r_0}{r_{\rm NS}}\right)^3 \sin\theta_m \sin(\Omega t) \, ,
\end{align} 
where  $B_0$ is the magnetic field strength at the surface of the neutron star (which is at located at $r_0$), and $\theta_m$ is the misalignment angle between $\vec{B}_{\rm NS}$ and $\vec {\Omega}$. The rotation vector of the neutron star is given by $\mathbf{\Omega}=\Omega{\bf \hat z_{\rm NS}}$ with $\Omega = 2\pi/T$ and $T$ the neutron star spin period.  We denote the polar angle of a given location with respect to the rotation axis $z_{\rm NS}$ to be $\theta_{\rm NS}$. The charge density is given by
\begin{align}
    n_{\rm GJ}(\vec{r}_{\rm NS}) = \frac{2 \mathbf{\Omega} \cdot \mathbf{B}_{\rm NS}}{e}\frac{1}{1 - \Omega^2 r^2 \sin^2 \theta_{\rm NS}} \, .
\end{align}
While we consider all charges to be electrons for simplicity, such that $n_{\rm GJ}$ simply gives the number of electrons, our conversion calculation still applies to more complicated plasmas. The plasma frequency of an electron plasma is $\omega_p \simeq \sqrt{4 \pi \alpha n_e / m_e}$, giving
\begin{equation}
    \omega_p(\vec{r}_{\rm NS}) = \sqrt{e {\bf B}_{\rm NS}\cdot {\bf \hat z}_{\rm NS}\frac{4\pi}{m_eT}\frac{1}{1 - \Omega^2 r^2 \sin^2 \theta_{\rm NS}}}\,,
\end{equation}
with
\begin{equation}
   {\bf B}_{\rm NS}\cdot {\bf \hat z}_{\rm NS}= \frac{B_0}{2} \left( \frac{r_0}{r_{\rm NS}} \right)^3 \big[3 \cos \theta_{\rm NS} \, {\bf \hat m} \cdot {\bf \hat r}_{\rm NS} - \cos \theta_m \big] \,,
\end{equation}
Similarly to Ref.~\cite{Hook:2018iia} we have also defined 
\begin{equation}
    {\bf \hat m} \cdot {\bf \hat r}_{\rm NS} = \cos \theta_m \cos \theta_{\rm NS} + \sin \theta_m \sin \theta_{\rm NS} \cos(\Omega t)\,.
\end{equation}

It should be noted that it is unlikely that the GJ model holds over the entirety (or any) of the neutron star. The nature of neutron star atmospheres is highly uncertain, due to both the difficulty in modeling such large electromagnetic systems and the lack of observational data~\cite{Kalapotharakos:2011vg,Cerutti:2016ttn,Philippov:2017ikm,Baring:2020zke}. Asides from possible higher multipole components of the magnetic field \cite{Bilous_2019}, pair creation cascades at the poles may lead to significantly higher electron densities, as well as highly boosted electrons \cite{RevModPhys.38.626,Harding_2006,Philippov:2017ikm}. Our formalism allows for both $\gamma\gg 1$ and $n_e\neq n_{\rm GJ}$, however, for specific examples we will remain within the GJ model.

\subsection{Directionality of emitted photons} \label{sec:results_direction}
To see the impact of a 3D formalism, we can take the simplest example and ask: how does the flux transfer change for different incoming axion trajectories? As we are considering the GJ model, for all cases $\gamma=1$, meaning that $\bar \omega_p=\omega_p$. In section~\ref{probability} we derived the flux transfer $R$ in a frame centered around the axion's velocity ($\bf \hat z$) and the direction of the $B$-field. However, as discussed above, the properties of the plasma are usually described in the frame of the neutron star, typically in spherical coordinates. To write down our $\bf \hat x$ and $\bf \hat y$ directions in terms of the incoming axion direction $\bf \hat z$ and magnetic field ${\bf\hat B}$ we can use
\begin{subequations}
\begin{align}
{\bf \hat x} &=\frac{1}{\sin\theta} {\bf\hat B}_{\rm NS} \times {\bf \hat z}, \\
    {\bf \hat y} &={\bf \hat z} {\bf \times \hat x}\,
    \end{align}
    \end{subequations}
\begin{figure*}[t!]
  \includegraphics[trim = 0mm 0mm 0mm 0mm, clip, width=0.45\textwidth]{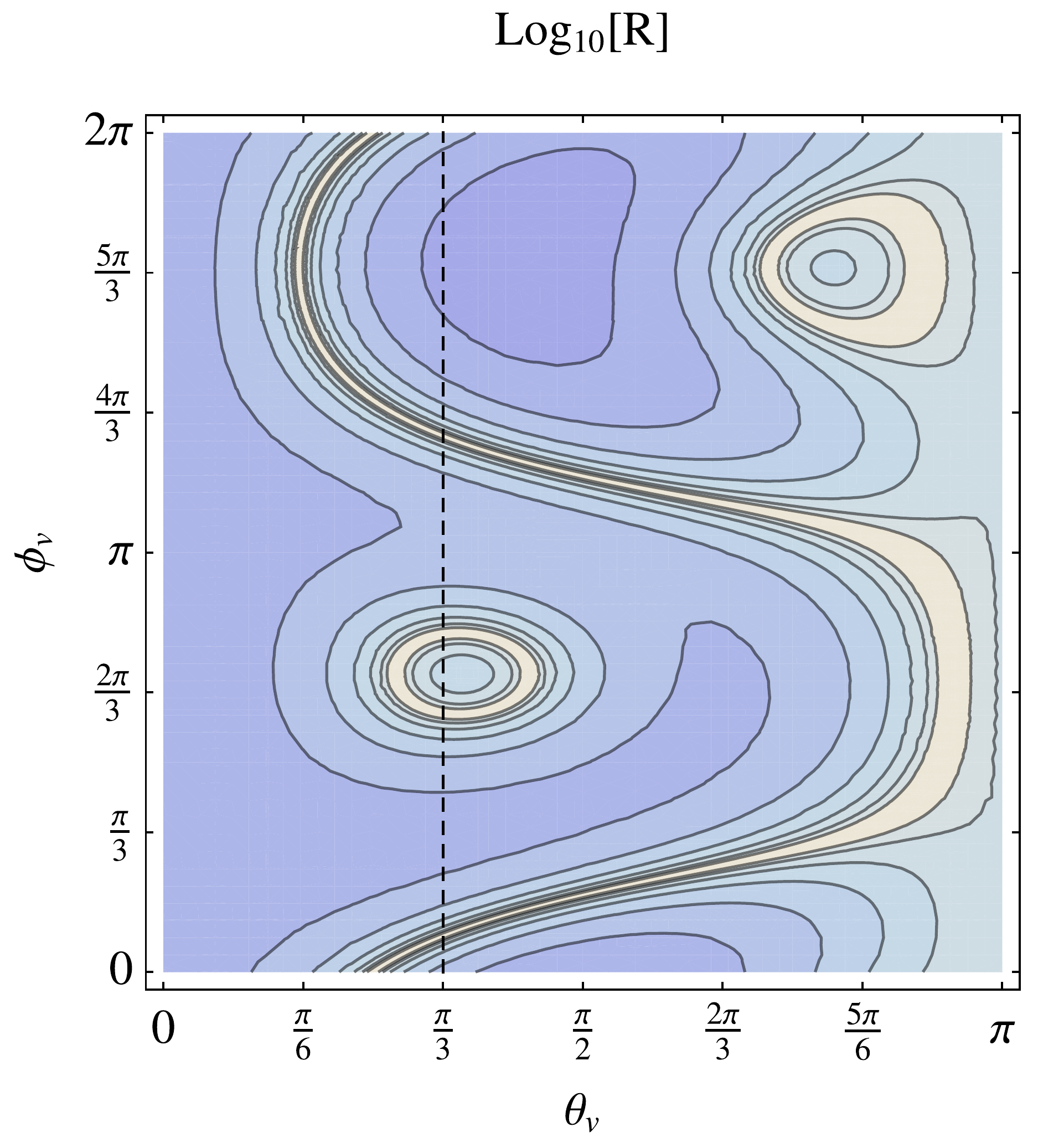}
  \includegraphics[trim = 0mm 0mm 0mm 0mm, clip, width=0.45\textwidth]{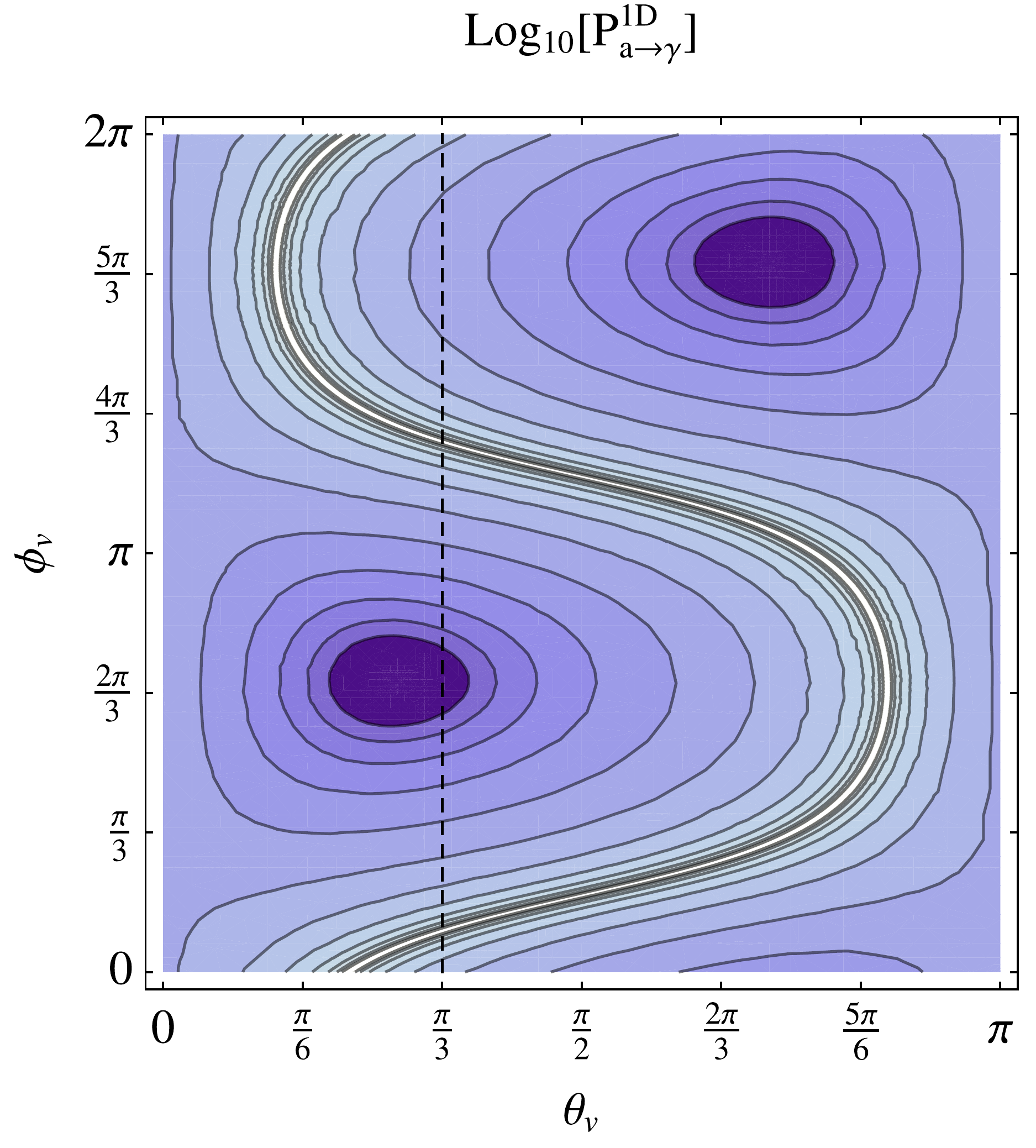}
\includegraphics[trim = 0mm 0mm 0mm 0mm, clip, width=0.08\textwidth]{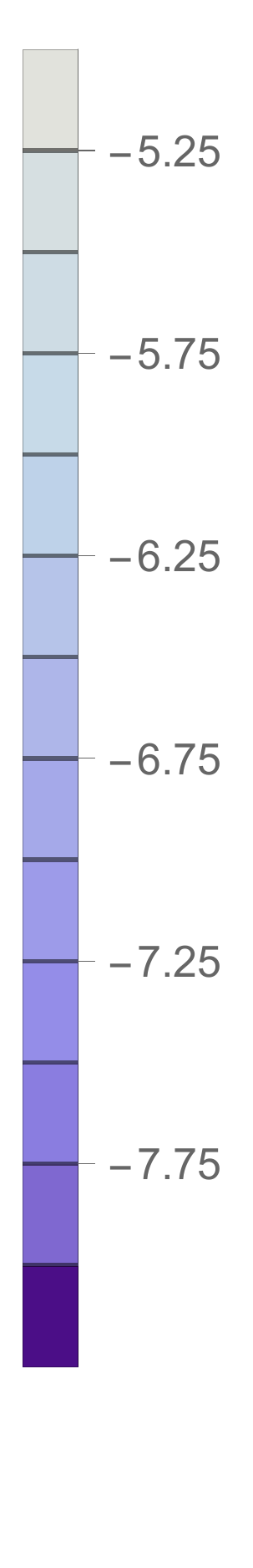}
\caption{{\bf Left panel:} Flux transfer $R$ as a function of incoming axion angles $\theta_v,\phi_v$ for the full 3D calculation. The neutron star has $B_{0}=10^{14}~$G, $\theta_m=0.4$ and $T=1$\,s, and the ratio is calculated at the conversion zone located at $\theta_{\rm NS}=0.5,\phi_{\rm NS}=0.6 \pi$ in the neutron star frame.  We assume an axion mass of 25$\,\mu$eV with a coupling strength $g_{a\gamma}=10^{-14}\,{\rm GeV}^{-1}$.  {\bf Right panel:} `Probability' of axion to photon conversion $P_{a\to\gamma}^{\rm 1D}$ as a function of incoming axion angles $\theta_v,\phi_v$ for a 1D calculation. The same neutron star and axion values are used. The black dashed line indicates the slice $\theta_v=\pi/3$ that is the focus of Fig.~\ref{fig:slicev}.}
 	\label{fig:point}
\end{figure*}
For an example, we consider a neutron star with $B_{0}=10^{14}~$G, $\theta_m=0.4$ and $T=1$\,s. Taking some point on the sphere, $\theta_{\rm NS}=0.5,\phi_{\rm NS}=0.6 \pi$, we can compare the flux transfer at the axion conversion surface as a function of the incoming axion angles, denoted by $\theta_v,\phi_v$ in the neutron star coordinate frame. Due to the dependence of the LO mode dispersion relation on $\theta$ and $k$, Eq.~\eqref{eq:resonancecond} indicates that the radius of conversion actually is slightly different for different incoming axion angles. However, the change in flux transfer due to small changes in conversion radius is subdominant compared to the impact of including aniosotropic effects of the medium.  For a benchmark, we take the axion to have mass $m_a=25\,\mu$eV and coupling $g_{a\gamma}=10^{-14}\,{\rm GeV}^{-1}$. Raising or lowering $m_a$ changes where the conversion zone is located: higher mass axions convert closer to the neutron star where magnetic fields are larger, usually leading to greater flux transfers. If the axion mass is too large, however, it is greater than the plasma frequency everywhere outside the surface of the neutron star and no resonant conversion can occur. As the plasma frequency is set by the magnetic field strength of the neutron star, for each star, there will be a maximum $m_a$ which will allow for resonant axion-photon conversion.

We plot the results of the 3D calculation, [the relativistic version of Eq.~\eqref{eq:prob1}], and of the 1D calculation, Eq.~\eqref{eq:prob1d}, in the right and left panels of Fig.~\ref{fig:point}, respectively.  To avoid divergences when the WKB approximation breaks down, we use a simple cut-off estimated by the typical lengths scales over which the second order derivatives should come into play, derived in Appendix~\ref{cut-off}
\begin{equation}
    R_{\rm cut{\mbox -}off}= g_{a\gamma}^2B_{\rm NS}^2\left(\frac{\pi r_c^2}{\bar\omega_pv_c^2}\right)^{2/3}\,.\label{eq:cut-offm}
\end{equation}
 We see that the number of photons generated on a given trajectory can differ between the 3D and the 1D formalism by several orders of magnitude. Furthermore, regions that were highly disfavoured in the 1D calculation can turn out to have maximal photon production when the aniosotropic nature of the medium is taken into account. 

\begin{figure*}
\centering
  \includegraphics[trim = 0mm 0mm 0mm 0mm, clip, width=.65\textwidth]{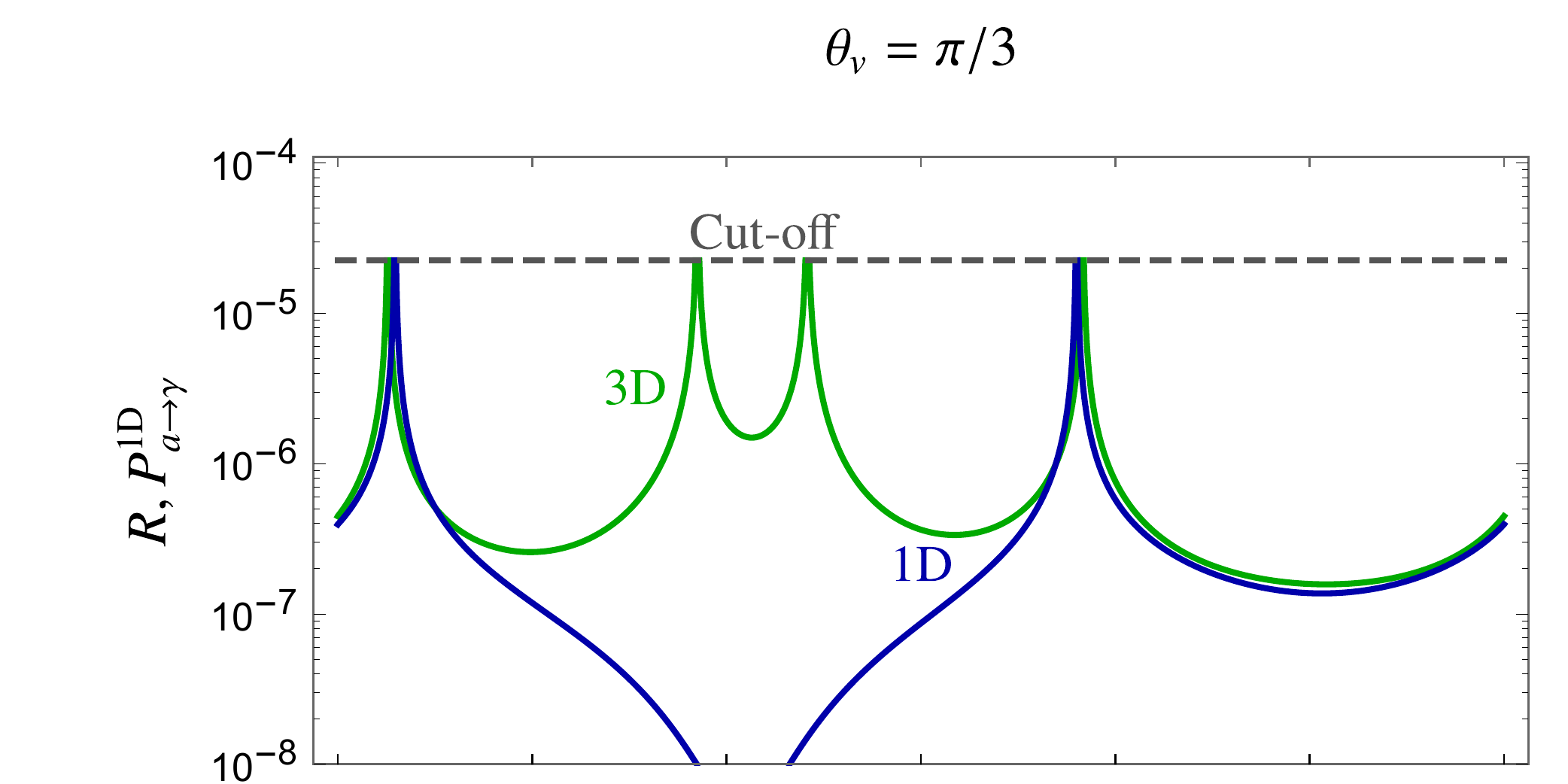} 
    \includegraphics[trim = 0mm 0mm 0mm 0mm, clip, width=.65\textwidth]{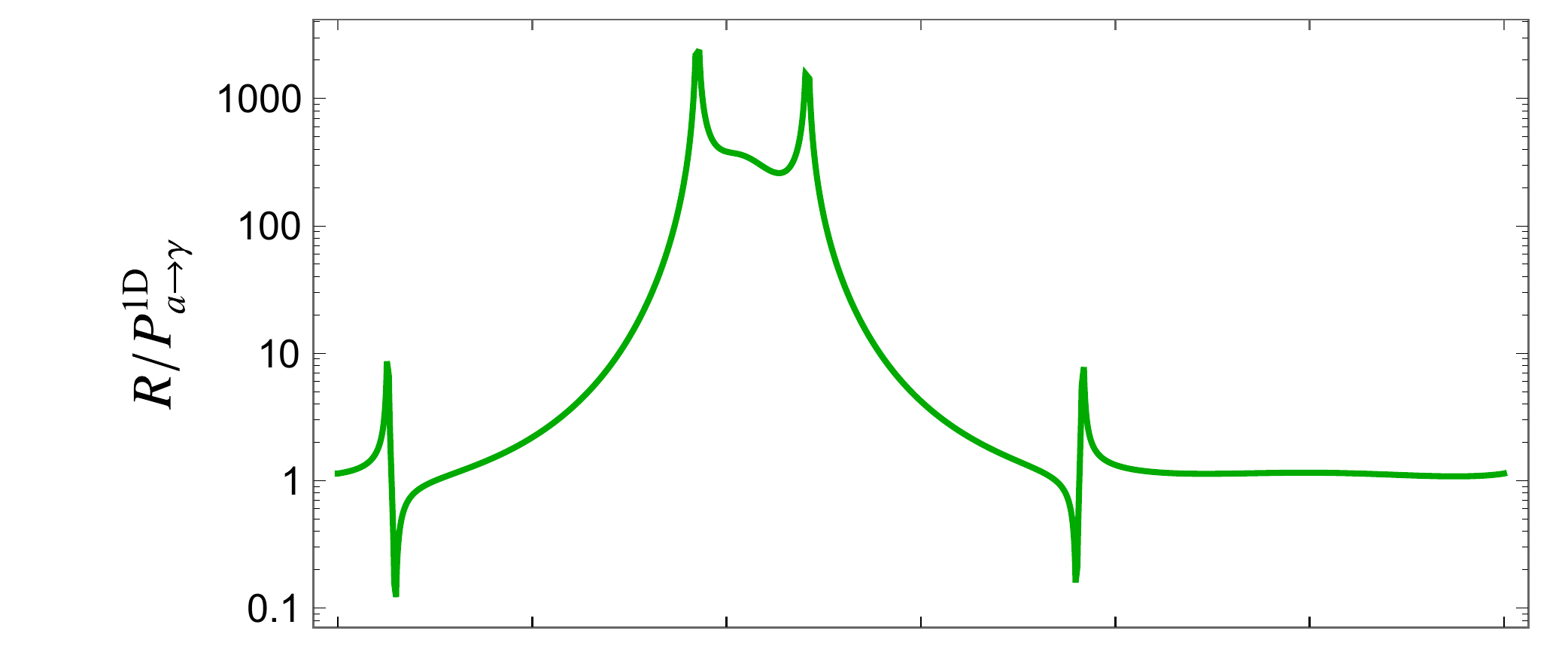} 
    \includegraphics[trim = 0mm 0mm 0mm 0mm, clip, width=.65\textwidth]{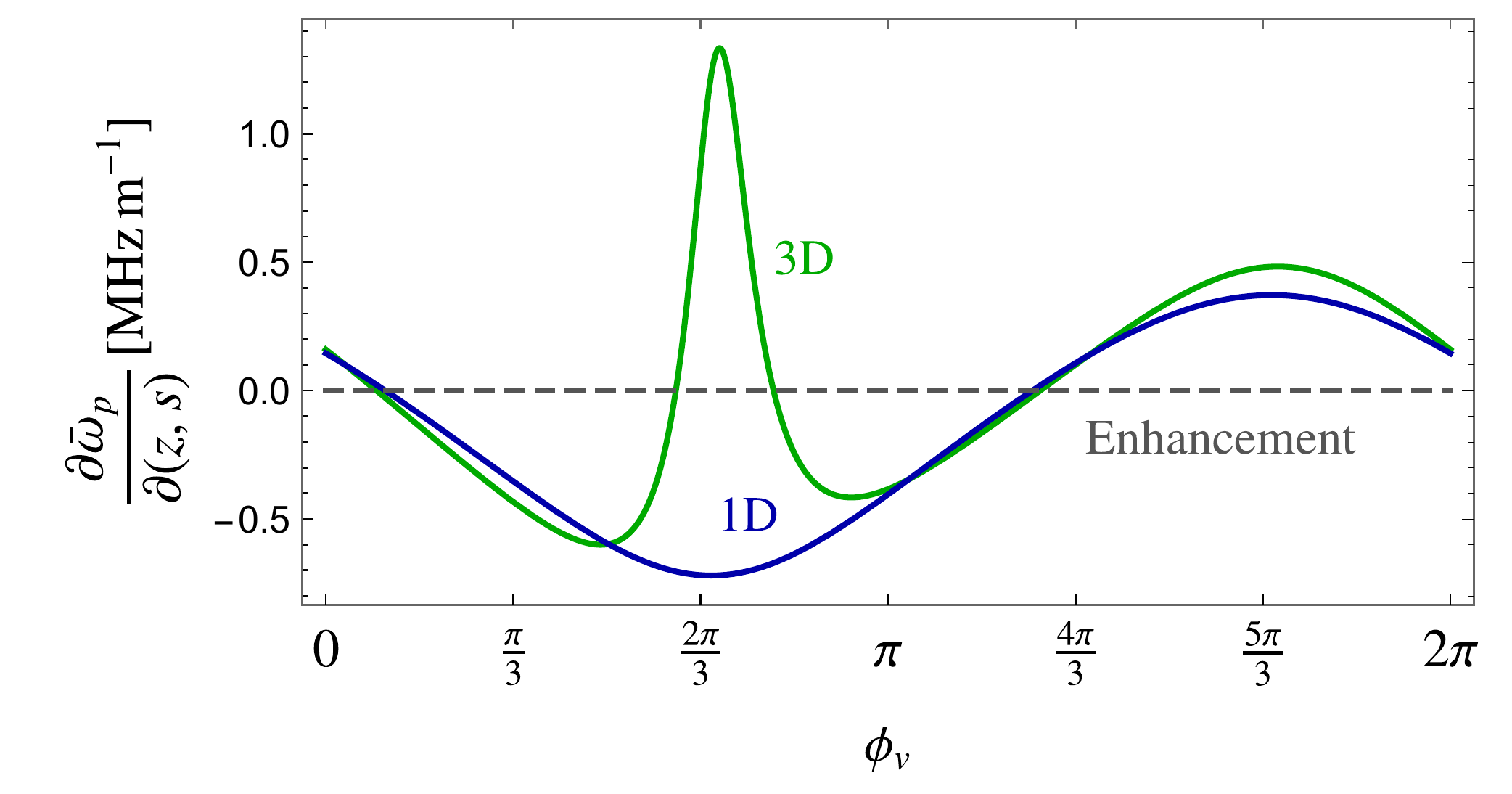} 
\caption{{\bf Top panel:} Axion-photon flux transfer $R$ as a function of incoming axion azimuthal angle $\phi_v$ for the full 3D calculation (green) and 1D calculation (blue). For the 1D calculation we use the `conversion probability' instead of the flux transfer. {\bf Middle panel:} Ratio of the full 3D axion-flux transfer $R$ to the 1D conversion probability. The flux is enhanced by up to three orders of magnitude with resepect to the 1D calcuation. {\bf Bottom panel:} Rate of change of the plasma frequency $\omega_p$ with respect to $s$ (green) and $z$ (blue). Axion photon conversion diverges when $\partial \omega_p/\partial (z,s)=0$, the gray dashed line. The neutron star has $B_{0}=10^{14}~$G, $\theta_m=0.4$ and $T=1$\,s, and the quantities are calculated at the conversion zone located at $\theta_{\rm NS}=0.5,\phi_{\rm NS}=0.6 \pi$ in the neutron star frame. We assume an axion mass of 25$\,\mu$eV with a coupling strength $g_{a\gamma}=10^{-14}\,{\rm GeV}^{-1}$ and incoming polar angle $\theta_v=\pi/3$.}
 	\label{fig:slicev}
\end{figure*}

To understand better the underlying physics behind the regions of enhanced and suppressed conversion, we can take a single slice of $\theta_v, \phi_v$ space. In Fig.~\ref{fig:slicev}, we show a slice of constant $\theta_v$, in particular, $\theta_v=\pi/3$. The top panel of Fig.~\ref{fig:slicev} shows the `conversion probability' for the 1D calculation and the flux transfer calculated with 3D effects, with the regulating cut-off flux transfer [Eq.~\eqref{eq:cut-offm}] shown by the gray dashed line. In the middle panel, we show the ratio $R/P_{a\to\gamma}^{\rm 1D}$: up to three orders of magnitude enhancement can occur for specific axion trajectories. The bottom panel shows the relevant derivatives for each calculation, $\partial \bar \omega_p/\partial z$ in the 1D case and $\partial \bar \omega_p/\partial s$ for the 3D calculation. The order-of-magnitude discrepancies between the 1D and 3D calculations of the flux transfer occur when $\partial \bar \omega_p/\partial z\neq \partial \bar \omega_p/\partial s$. While there are still differences caused by correctly calculating the energy stored in the LO mode and the correct angular dependence, the primary difference between calculations comes from the difference between $ \bf \hat s$ and $ \bf \hat z$: when $\partial \omega_p/\partial s\simeq \partial \omega_p/\partial z$, the answers are similar up to the different angular dependence. The divergences in $R,P_{a\to\gamma}$ (which are regulated by the cut-off) occur when $\partial \omega_p/\partial (s,z)=0$, as indicated by the gray dashed line in the bottom panel of Fig.~\ref{fig:slicev}. In these examples, $\theta'$ is negligible and so the cut-off is a stronger regulator. Note that the relationship between $\bf \hat s$ and $\bf \hat z$ does not depend on $m_a$: even for very non-relativistic axions which could convert far out from the NS there would still be significant differences in the flux transfer in a given direction.

We can see that treating the LO mode as essentially a transverse mode (possibly with different dispersion, as in~\cite{Witte:2021arp}) leads to a significant miscalculation of the axion-photon flux transfer. To have a reliable calculation of the axion-photon flux transfer for a given axion trajectory, one must take into account both the full energy stored in the LO mode, and, more importantly, the distinction between the direction of the axion's momentum, $\bf \hat z$, and the direction over which the energy density evolves, $\bf \hat s$. 

\subsection{Integrated power} \label{sec:results_power}
Asides from the directional dependence of the photons emitted from a given point, one may also be concerned with the total power emitted over some portion of the neutron star. Studies of ray-tracing suggest that the photons bend relatively quickly as plasma frequency drops~\cite{Witte:2021arp,Battye:2021xvt}. The momentum acquired by the photon as it transitions to vacuum must come from the gradient of the medium, so photons that begin mildly relativistic would become approximately parallel to the change in $\bar \omega_p$. Because of this, one can make a rough estimate that the total power exiting the neutron star will end up on the same trajectory, determined by $\nabla\bar\omega_p$. We stress however that the axion can be reasonably relativistic, and so one should do a full ray-tracing calculation for reliable results. 

To calculate the power radiated from a given point on the neutron star ${\vec r}_{\rm NS}$, we must integrate over the incoming axion angles $\theta_v,\phi_v$, weighted with an appropriate phase space distribution. To focus on the difference caused by properly treating the aniostropy of the medium, we will neglect the small differences in conversion radii for different axion trajectories and use a single radius of conversion given by the transverse photon dispersion, $m_a=\omega_p$. The variation due to choosing different radii seems to be at most a factor of 2 when one considers the longitudinal photon dispersion ($\omega=\omega_p$) instead, and does not change the general relationship between the two calculations. Similarly, we neglect contributions from $\theta'$, which again gives a subdominant contribution for the cases we consider. The distribution-weighted axion-photon flux transfer $R_W$ is given by
\begin{equation}
    R_W(\vec{r}_{\rm NS}) = \int d\Omega_v f_5(\vec{r}_{\rm NS},\theta_v,\phi_v ) R(\vec{r}_{\rm NS},\theta_v,\phi_v ) \,,
\end{equation}
where $f_5$ is the five dimensional phase space defined in Eq.~\eqref{eq:fivespace}. For the 1D case we use the 1D `conversion probability', Eq.~\eqref{eq:prob1d} instead of $R$.
\begin{figure*}
\centering
  \includegraphics[trim = 0mm 0mm 0mm 0mm, clip, width=.65\textwidth]{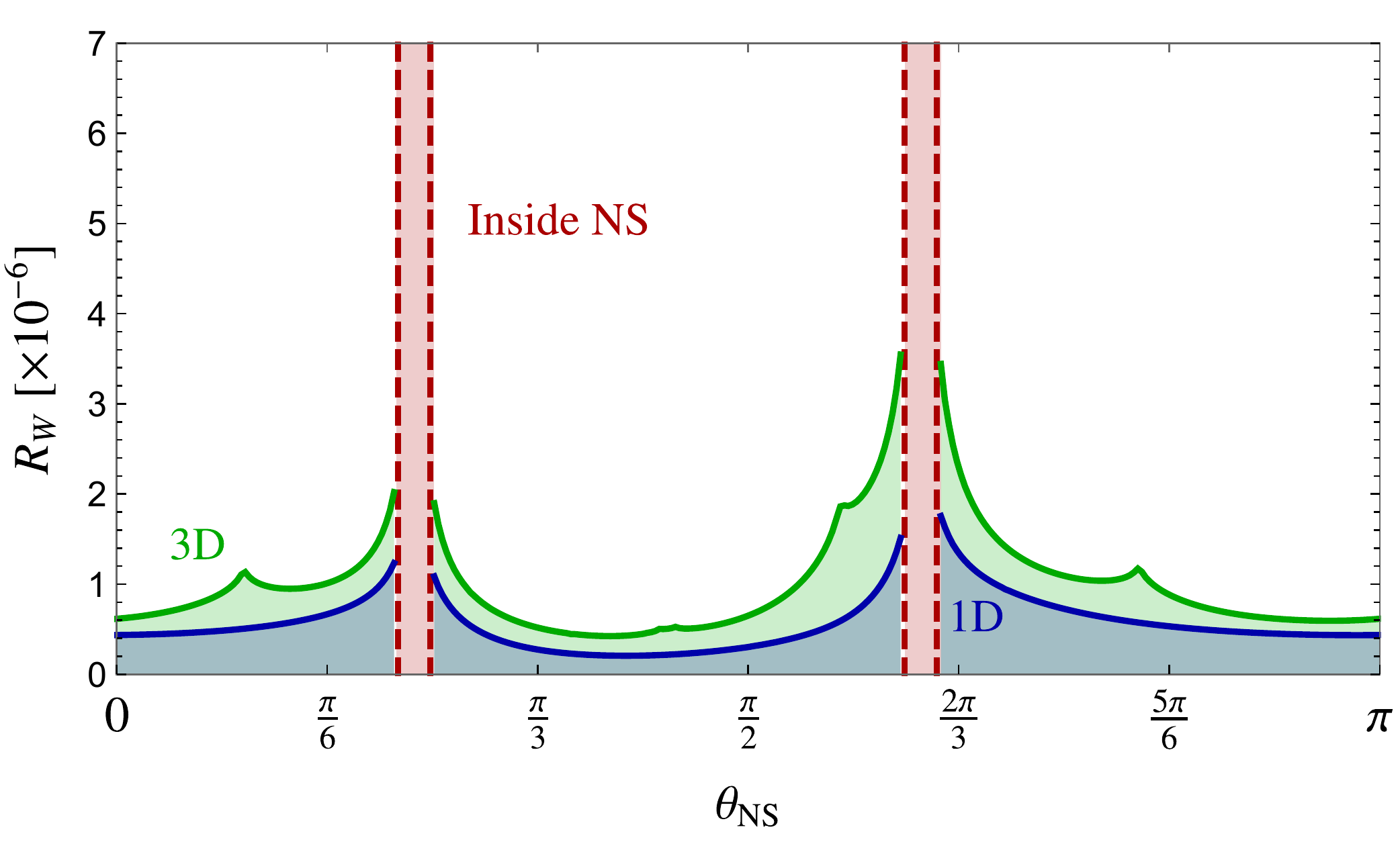} 
\caption{Axion photon conversion integrated over axion phase space, $R_{W}$, as a function of incoming neutron star polar angle $\theta_{\rm NS}$ for 3D (green) and 1D (blue) calculations. The 3D calculation uses the full flux transfer of axions to photons, whereas the 1D case only considers the `conversion probability' interpreted from a `Schr\"odinger-like' equation. We choose a neutron star with $B_{0}=10^{14}~$G, $\theta_m=0.4$ and $T=1$\,s. The functions are calculated at the conversion zone located at an azimuthal angle $\phi_{\rm NS}=\pi$ in the neutron star frame. We assume an axion mass of 25$\,\mu$eV with a coupling strength $g_{a\gamma}=10^{-14}\,{\rm GeV}^{-1}$ as well as an isotropic velocity distribution of the axions. We mark the region where the conversion radius would be less than the neutron star radius $r_0$ with the red dashed lines labeled `Inside NS'.}
 	\label{fig:sliceN}
\end{figure*}

We can now show $R_W$ for a slice along the neutron star conversion surface assuming an isotropic axion distribution. We keep $\phi_{\rm NS}$ fixed at $ \pi$ and vary $\theta_{\rm NS}$ to derive Fig.~\ref{fig:sliceN}. The structure generated in a slice through $\theta_{\rm NS}$ is in general behaviour independent of the specifically chosen $\phi_{\rm NS}$.  Integrating over all photon directions generally smooths out differences between the 3D and 1D calculations. However, there are still noticeable differences. Overall, the 3D calculation leads to more energy stored in the EM fields, leading to more photons radiated from the neutron star. As the 1D calculation neglected energy stored in the non-transverse components of the Hamiltonian, we would naturally expect to see additional `conversion' just from correctly counting the number of photons generated. This enhancement is particularly noticeable around the neutron star `throat', i.e. the region where the conversion radius vanishes as ${\bf\Omega}\cdot{\bf B}_{\rm NS}\to 0$, just outside the red boarders in Fig.~\ref{fig:sliceN}. Inside the neutron star itself, the plasma density no longer follows the GJ model; given the extreme densities inside a neutron star the plasma frequency is much larger than the axion masses of interest everywhere inside a neutron star. Thus, no propagating photons can be produced from axion-photon conversion inside a neutron star. 
One would expect that issues of photon curvature, as explored in Ref.~\cite{Witte:2021arp}, as well any concerns about gravity bending the axion path would be most pronounced in the throats, making this region generally unreliable. One would likely need to do numerical calculations for a robust solution including a gravitationally bending axion trajectory. 

Having a similar total conversion rate for isotropic dark matter infall is not surprising, as ultimately the length scales and magnetic fields that give the general order of conversion are the same. However, where the axions are converted, and in what directions the photons emerge, is significantly impacted. The plasma effectively lenses the outgoing photons, which when combined with the limited line of sight and Doppler shifting of light can give a complicated, time dependent signal. Thus any aspect which changes the location and direction of conversion will impact axion searches significantly. Because of this we focus on the changes to the power and directionality of a given point on the neutron star, as the total power radiated from a neutron star is not usually relevant to an individual observation of an individual neutron star. In addition, for compact clumps of axions, such as axion stars~\cite{Kaup:1968zz,Ruffini:1969qy,Tkachev:1991ka,Chavanis:2011zi,Visinelli:2017ooc,Prabhu:2020yif} or miniclusters~\cite{Hogan:1988mp,Kolb:1993zz,Kolb:1993hw,Kolb:1994fi,Eggemeier:2019khm,Edwards:2020afl}, one would expect a much stronger dependence in the resulting signal depending at what angle the compact object hits the neutron star. Thus, changes to the production location and direction can have a disproportionate effect on observational constraints. 

\section{Conclusions} \label{sec:conclusions}
The axion is one of the premier candidates for dark matter. It displays unique wavelike phenomenology requiring both novel techniques in the lab and astrophysical searches. Resonant conversion to photons in the atmospheres of neutron stars is one such possible method for discovering the axion. Neutron stars are complicated systems and therefore calculating the projected signal is extremely challenging. Here we have focused on one piece of the larger puzzle, and together with recent advances in ray-tracing help lay the road to a robust prediction for what might be observed on Earth.

To this end, we have calculated the production of photons via axions in strongly magnetised, anisotropic environments. The prototypical example of such an environment is a neutron star, but our calculation is generic. Previous calculations have not explored the anisotropic nature of the medium and have focused only on the transverse part of photon modes. We show, however, that the mixing of transverse and longitudinal modes into Langmuir-O modes causes the direction over which the amplitude of the electromagnetic fields evolve to be different from the direction of the axion. We find that rather than evolving in the direction of propagation, the waves evolve also in the transverse direction. This directional difference can lead to orders of magnitude change in the flux transfer $R$ in a given direction.

Furthermore, taking only the transverse component of the stored energy into account neglects a significant part of the stored energy, or in other words, miscounts the number of photons that are converted. Taking only the transverse component essentially confuses a classical, Schr\"odinger-like solution with the quantum mechanical conversion probability. We show that taking both features into account significantly modifies the probability of conversion over a given axion-trajectory. While a truly complete solution would require numerically solving the equations of motion in a varying medium, we provide simple analytical formulae which can be used to estimate the radio signals generated in neutron stars by axions. This will allow the robustness of axion limits to be increased in the case of non-observation, such as Refs.~\cite{Foster:2020pgt,Darling:2020uyo,Darling:2020plz,Battye:2021yue}, or allow one to estimate the neutron star and axion properties in the case of a discovery. 

\section*{Acknowledgements}
The authors thank Alice Harding, Wynn Ho and Anna Watts for helpful discussions on neutron star magnetospheres. The authors also thank Thomas Edwards and Sam Witte for helpful discussions, and Dion Noordhuis for pointing out a sign error. AM, DM and ML are supported by the European Research Council under Grant No. 742104 and by the Swedish Research Council (VR) under Dnr 2019-02337 “Detecting Axion Dark Matter In The Sky And In The Lab (AxionDM)". DM is further supported by the Swedish Science Council through grant 2018-03641.
SB acknowledges support by NSF Grant PHY-1720397, DOE HEP QuantISED award \#100495, the Gordon and Betty Moore Foundation Grant GBMF7946, and Simons Investigator Award 824870.

\appendix

\section{Longitudinal component}\label{longitudinal}
In the main text, we solve the axion-Maxwell equations by eliminating the longitudinal terms from the equations. However, this does not mean that the longitudinal modes do not exist. To solve for the longitudinal component excited by the axion, we can rewrite Eq.~\eqref{eq:mixed21} to find 
\begin{equation}
    E_z=\frac{\frac{\partial^2 E_y}{\partial z\partial y}-\bar\omega_p^2\cos\theta\sin\theta E_y-\omega^2 g_{a\gamma} a B_{\rm NS}\cos\theta}{\omega^2-\bar\omega_p^2\cos^2\theta} \label{eq:long} \, .
\end{equation}
First we should consider the case without an axion to find the normal electromagnetic modes of the system. In the absence of the axion, for a homogeneous medium one finds that
\begin{equation}
	  E_z=\frac{\frac{\partial^2 E_y}{\partial z\partial y}-\bar\omega_p^2\cos\theta\sin\theta}{\omega^2-\bar\omega_p^2\cos^2\theta} E_y\,,
\end{equation}
allowing us to define the propagating Langmuir-O (LO) mode, which becomes an ordinary transverse wave when $\theta=\pi/2$ and longitudinal when $\theta = 0$ in the non-relativistic limit~\cite{PhysRevE.57.3399},
\begin{equation}
	    {\bf\hat E}_{\rm LO}=\frac{-1}{\sqrt{1+\frac{\bar\omega_p^4\cos^2\theta\sin^2\theta}{(\omega^2-\bar\omega_p^2\cos^2\theta)^2}}}\left({\bf \hat y}-\frac{\bar\omega_p^2\cos\theta\sin\theta}{\omega^2-\bar\omega_p^2\cos^2\theta}{\bf \hat z}\right )\,,
	\end{equation}
	where we have inserted a factor of $-1$ so that ${\bf\hat  E}_{\rm LO}={\bf\hat  B}_{\rm NS}$ in the non-relativistic limit.
Including the axion current, and knowing that we will solve for $E_y$ directly, we can neglect the subdominant derivative terms to find
\begin{equation}
	\tilde E_z\simeq-\frac{\bar\omega_p^2\cos\theta\sin\theta}{\omega^2-\bar\omega_p^2\cos^2\theta}\tilde E_y-\frac{\omega^2\cos\theta}{\omega^2-\bar\omega_p^2\cos^2\theta}g_{a\gamma}B_{\rm NS}\tilde a\,.\label{eq:long2}
\end{equation}
 The first term is just the longitudinal component of the LO mode, with the second corresponding to direct axion mixing. In the limit of a perfectly aligned magnetic field with $\theta=\{0,\pi\}$ Eq.~\eqref{eq:long2} reduces to the expression for a pure longitudinal excitation in a lossless plasma~\cite{Millar:2017eoc,Lawson:2019brd}, 
\begin{equation}
    \tilde E_z=-\frac{g_{a\gamma}B_{\rm NS}\tilde a}{1-\frac{\bar\omega_p^2}{\omega^2}}\,.
\end{equation}
We can see that the two different terms in $E_z$ have very different propagation behaviour. After some distance (given by the conversion length $L$) $\tilde E_y$ will be out of phase with $\tilde a$. Notice that the terms proportional to $\tilde a$ are not generally propagating: as soon as the axion mixing goes away, so to does this contribution to $E_z$. 

Thus, for the purpose of solving for an outgoing radio signal from the neutron star, the mode that the axion couples to is the LO mode, which will adiabatically transform into a transverse mode as it exits the system~\cite{PhysRevE.57.3399}.

\section{Stationary phase approximation}
\label{stationaryphase}
\label{sec:statphase}
To provide an analytic expression for axion-photon conversion, in Eq.~\eqref{eq:schro} in the main text we derived a Schr\"odinger-like equation
\begin{align}
    i\frac{\partial \tilde E_y}{\partial s}=\frac{1}{2k}\left (m_a^2-\xi\bar\omega_p^2-ikD\right )\tilde E_y -\frac{1}{2k}\frac{\omega^2 \xi}{\sin\theta} g_{a\gamma} \tilde a B_{\rm NS}\,.
\end{align}
which has solution
\begin{align}
 i E_y(s)=-\int_0^sds'\frac{1}{2k}\frac{\omega^2 \xi}{\sin\theta} g_{a\gamma} \tilde a B_{\rm NS} e^{-\int_0^{s'}ds''D/2}e^{-i\int_0^{s'}ds''\frac{1}{2k}\left (m_a^2-\xi\bar\omega_p^2\right )}\,. \label{eq:shrosol}
\end{align}
Rather than a numeric solution, we can use the stationary phase approximation to find an analytic solution. The premise of the stationary phase approximation is that highly oscillatory integrands tend to cancel when integrated over many oscillation periods, and that the dominant contribution therefore often comes from stationary points, say $s_0$, where the phase, $f(s)$, is locally constant, i.e.~$f'(s_0)=0$. In the standard form of the stationary phase approximation, the non-oscillatory factor of the integrand, say $g(s)$, is approximated as constant around $s_0$, and the phase is expanded to quadratic order: $f(s) \approx f(s_0) + \tfrac{1}{2} f''(s_0) (s-s_0)^2$, and the limits of integration are taken to $\pm\infty$. For $s>s_0$, this gives the approximation, 
\begin{equation}
    \int_0^{s} ds\, g(s)e^{if(s)}\approx 
     g(s_0)e^{if(s_0)} \int_{-\infty}^{\infty}ds\, e^{\frac{i f''(s_0)}{2}  (s-s_0)^2}
    =
    g(s_0)e^{if(s_0)+{\rm sign}[f''(s_0)]i\pi/4}\sqrt{\frac{2\pi}{|f''(s_0)|}}\,.
    \label{eq:statphase}
\end{equation}

The  quality of the stationary phase approximation of course depends on the accuracy of the assumptions that it relies on. It's convenient to define a (squared) `conversion length' as
   \begin{equation}
    L^2=\frac{\pi}{|f''(s_0)|}\,.
\end{equation}
In  the limit where $\sqrt{|f''(s_0)|}$ is larger than all other scales in the problem, the conversion length is small, and  the stationary phase approximation is expected to be very accurate. However, in many physical situations there are no large hierarchies of scales, and one needs to consider corrections to the standard stationary phase formula. Moreover, extending the integration range to $\pm \infty$ is not always justified, and the full integral that one wants to evaluate may only be well-defined in a small region around the stationary point, such as for our Eq.~\eqref{eq:shrosol}. 

In the context of  relativistic axion-photon conversion, these issues were recently discussed in Ref.~\cite{Marsh:2021ajy}, where corrections to Eq.~\eqref{eq:statphase} were derived from considering a finite integration range, and accounting for contributions neglected in the standard stationary phase approximation. For the purpose of our analysis, the main point is that when one considers the integration range $[s_0 - \Delta s, s_0 + \Delta s]$, the stationary phase approximation gives     
\begin{align}
    \int_0^{s} ds\, g(s)e^{if(s)}&\approx 
     g(s_0)e^{if(s_0)} \int_{s_0 - \Delta s}^{s_0 + \Delta s}ds\, e^{\frac{i f''(s_0)}{2}  (s-s_0)^2}
     \\
    &=
    g(s_0)e^{if(s_0)+{\rm sign}[f''(s_0)]i\pi/4}\sqrt{2} L\, {\rm Err}\left(\sqrt{-i\pi/2} \frac{\Delta s}{L} \right)\, ,
    \label{eq:statphasecorr}
\end{align}
where ${\rm Err}$ denotes the error function. The norm of the error function grows linearly with $\frac{\Delta s}{L}$ for small arguments, and performs slowly damped oscillations around unity for $\frac{\Delta s}{L}>1$. This in particular implies that the standard formula only applies when it is justified to take $\Delta s \gg L$. Equation \eqref{eq:statphasecorr} applies for any ratio of $\Delta s/L$, but simplifies considerably when $\Delta z <L$, and the integral becomes
\begin{equation}
\Big|\int_0^{s} ds\, g(s)e^{if(s)} \Big| = 2 |g(s_0)|  \Delta z\, ,
\end{equation}
up to corrections to cubic order in $\Delta z/L$. This implies that the formula \eqref{eq:statphase} can be regulated and modified in a very simple way, by replacing $L \to \sqrt{2 }\Delta z$, when it is justified to consider $\Delta z < L$, for example when the WKB approximation breaks down, or the assumption of constant $\theta$ fails. For a more detailed discussion of how to evaluate also the contributions from $|z|>\Delta z$, see Ref.~\cite{Marsh:2021ajy}.

The most important factor comes from $f''(s_0)$. While we have so far considered $\theta$ to be constant, if it is slowly changing the largest impact will be felt in the conversion length~\cite{Witte:2021arp} as the impact of dephasing the axion and photon is much greater than the small changes to the definition of $\bf \hat y$. We can then see that
\begin{equation}
  \frac{\partial}{\partial s}  \left(\frac{1}{2k}\left [m_a^2-\xi\bar\omega_p^2\right ]\right)=-\frac{\bar\omega_p\bar\omega_p'\sin^2\theta}{k\left(1-\frac{\bar\omega_p^2}{\omega^2}\cos^2\theta\right)^2}-\frac{\bar\omega_p^2(\omega^2-\bar \omega_p^2)\theta'\sin\theta\cos\theta}{\omega^2k\left(1-\frac{\bar\omega_p^2}{\omega^2}\cos^2\theta\right)^2}\,.
\end{equation}

We now apply the standard stationary phase approximation of Eq.~\eqref{eq:statphase} to the electric field of Eq.~\eqref{eq:shrosol}, which gives 
\begin{equation}
    \tilde E_y[s_0+L/2]=\omega^2 \sqrt{\frac{\pi}{2k\left |\bar\omega_p\bar\omega_p'+\frac{\omega^2-\bar\omega_p^2}{\omega^2\tan\theta}\bar\omega_p^2\theta'\right |}}e^{\int_0^{s_0}ds'D/2}g_{a\gamma} \tilde a B_{\rm NS}\,,\label{eq:stationaryA}
\end{equation}
where we have neglected an overall phase and where $\bar\omega_p'=\partial \bar\omega_p/\partial s$. This is the analytical form of axion-induced $E_y$ that we will use in our subsequent analysis.  
The stationary point condition (cf.~$f'(s_0)=0$) translates into the `resonance condition':
\begin{equation}
	\bar\omega_p(s)^2=\frac{m_a^2\omega^2}{m_a^2\cos^2\theta+\omega^2\sin^2\theta}\,.
\end{equation}
In the non-relativistic limit, $\omega \to m_a$, and the resonance condition simply becomes $\bar\omega_p = m_a$. 
The conversion length evaluates to
\begin{equation}
    L= \frac{\sin \theta}{\xi} \sqrt{\frac{\pi k}{\left |\bar\omega_p\bar\omega_p'+\frac{\omega^2-\bar\omega_p^2}{\omega^2\tan\theta}\bar\omega_p^2\theta'\right |}}\,.
\end{equation}
For most axion trajectories, this conversion length is smaller than the length scales associated with the gravitational curvature of the axion trajectory, the curvature of the photon path due to the anisotropic medium, and the scale over which the magnetic field changes. When this is the case, the standard stationary phase approximation applies. However, this is not always the case, and in some cases the conversion length can get significantly enhanced, as we now discuss.

\section{Regulating divergences}
\label{cut-off}
The flux transfer $R$, given by Eq.~\eqref{eq:prob1}, was calculated assuming the WKB approximation holds: i.e., that the axion momentum is much larger than the first derivatives of $\tilde E_y$, which are larger than the second derivatives. However, $R$ diverges in the limit that the first derivative vanishes (i.e., the conversion length becomes infinite). Such a divergence must be regulated for two reasons. The first is the breakdown of the WKB approximation, and the second is the breakdown of the assumption that the conversion happens over length scales smaller than other relevant ones in the neutron star, for example the radius of photon bending due to the changing media.

In the first case, it is evidently untrue that the vanishing first derivatives are larger than the second derivatives. Rather than attempting to solve the full 3D equations by hand with higher derivatives, as $E_x,E_y$ and $E_z$ all appear with their second derivatives, we can use a simple estimate to institute a cut off on the diverging flux transfers. In the 1D case, it was shown in Ref.~\cite{Hook:2018iia} that an order-of-magnitude estimate could be obtained by using the simple formula
\begin{equation}
	R\simeq \frac{B_{\rm NS}^2g_{a\gamma}^2L^2}{v_c^2}\,,
\end{equation}
where $L$ is the conversion length. The assumption of coherent conversion requires that the change in phase between two locations is small. Thus, we can estimate $L$ by checking the distance over which $k$ changes. We can approximate $k$ changing over some length $L$ by writing
\begin{equation}
	k(L)=k_0+k_1L+k_2L^2\,,
\end{equation}
where destructive interference occurs when $\delta k=k-k_0$ satisfies $\delta k L\simeq \pi$ (as the phase enters via $e^{i(kL-\omega t)}$. As we are concerned with the case where $k_1L\lesssim k_2L^2$, we can neglect the linear term and see that 
\begin{equation}
	L^3\simeq \pi/k_2\,.
\end{equation} 
As the only change in $k$ comes from the changing plasma frequency $\bar\omega_p$, in the limit of vanishing first derivatives $kk_2=\bar\omega_p\bar\omega_{p,2}$, where similarly
\begin{equation}
	\bar\omega_p(L)\simeq \bar\omega_{p,0}+ \bar\omega_{p,2}L^2\,.
\end{equation} 
To get a conservative estimate for $L$, we can use that at second order, all directions may play a role. Thus while the second derivatives with respect to some directions might be small or vanishing, we know that the radial direction will have a power law dependence. Thus, assuming that the $\hat {\bf r}$ direction is the relevant one will give a fairly conservative estimate for the total conversion probability. We stress that this is not a full solution of the $3$D Maxwell equations, but rather a conservative way of regulating our WKB formalism. Taking the power law dependence of $\bar\omega_p$, we then find that
\begin{equation}
	L=\left(\frac{\pi v_cr_c^2}{\bar\omega_p}\right)^{1/3}\,.
\end{equation}
Putting in some typical numbers, we see that the typical conversion length should increase by about an order of magnitude, when compared to purely radial estimates of Ref.~\cite{Hook:2018iia}, leading to a hundredfold increase in the conversion probability. 

The second case relates to the overall trajectory of photons in the medium, for example, that the photon bends and so dephases with the axion, or other curvature effects of the medium~\cite{Witte:2021arp}. While such a constraint may provide a more stringent limit, particularly in the neutron star throat, to calculate such effects in full requires ray tracing calculations~\cite{Leroy:2019ghm,Witte:2021arp,Battye:2021xvt}, which are beyond the scope of this work. When these effects are considered, one should use the stricter of the two cut-offs. As we are only interested in demonstrating the influence of a 3D, aniosotropic calculation of the flux transfer rather than a start to finish signal calculation for observational purposes, we can simplify matters and just use a cut-off based on the second derivatives of the $E$-field,
\begin{equation}
    R_{\rm cut{\mbox -}off}= g_{a\gamma}^2B_{\rm NS}^2\left(\frac{\pi r_c^2}{\bar\omega_pv_c^2}\right)^{2/3}\,.\label{eq:cut-off}
\end{equation}

\providecommand{\href}[2]{#2}\begingroup\raggedright\endgroup

\end{document}